\documentclass[aps,prx,reprint,superscriptaddress,twocolumn]{revtex4-2}

\usepackage{amsmath,amsfonts,amsthm,amsbsy}

\usepackage{graphicx}
\usepackage{braket}

\usepackage[pdftex,colorlinks=true,linkcolor=blue,citecolor=blue,urlcolor=blue]{hyperref}

\usepackage{calligra}

\DeclareMathAlphabet{\mathcalligra}{T1}{calligra}{m}{n}
\newcommand{\chern}{\mathcalligra{C}}

\begin{document}

\title{Wave topology brought to the coast}

\author{A. Venaille}
\email{antoine.venaille@ens-lyon.fr}

\author{P. Delplace}
\email{pierre.delplace@ens-lyon.fr}

\affiliation{Univ Lyon, Ens de Lyon, Univ Claude Bernard, CNRS, Laboratoire de Physique, F-69342
Lyon
}

\date{\today}


\begin{abstract}
Since the pioneering work of Kelvin on Laplace tidal equations, a zoology of trapped waves have been found in  the context of coastal dynamics. Among them, the one originally computed by Kelvin plays a particular role, as it is an unidirectional mode filling a frequency gap between different wave bands. 
The existence of such Kelvin waves is robust to changes in the boundary shape and  in changes of the underlying model for the coast. This suggests a  topological interpretation that has yet up to now remained elusive. Here we rectify the situation, by taking advantage of a reformulation of the shallow water dynamics that highlights an analogy with the celebrated Haldane model in condensed matter physics.  For any profile of bottom topography, the number of modes that transit from one wave band to another in the dispersion relation is predicted by computing a first Chern number describing the topology of complex eigenmodes in a dual, simpler wave problem.
\end{abstract}

\maketitle

\section{Introduction}
 
Coastal Kelvin waves were derived in 1880 in an attempt to solve tidal Laplace equations in oceanic basins \cite{kelvin}. These equations model the linear dynamics of  surface shallow water in the presence of the Coriolis force.
Kelvin waves are trapped modes that have the noteworthy properties to fill a frequency gap between different wavebands and to travel as surface waves in a cyclonic manner along the coasts. Such waves are now routinely observed along continental margins or sufficiently large lakes. 

{\color{black} These remarkable features bear striking similarities with peculiar electronic states found over the last decades at the boundary of exotic materials called topological insulators, and more specifically Chern insulators. In such material, the number of unidirectional trapped waves along the boundary  
is ruled by a single number, the first Chern number, that describes the topology of normalized eigenmode bundles in an abstract dual \textit{bulk} problem. This bulk problem describe similar waves, but in unbounded geometries and with homogeneous coefficients in the linear operator. 
The first Chern number of the abstract bulk problem predicts the number of unidirectional boundary trapped modes in the original wave problem through a celebrated bulk-boundary correspondence \cite{hatsugai1993chern}. Such boundary trapped modes are often said to have a topological origin, or to be topologically protected.}

{\color{black} Topology offers a powerful tool to make predictions on a complicated problem without having to solve this problem. 
Here, robust properties of  partial differential equations with inhomogeneous coefficients can be deduced by looking at the topological properties of the much simpler dual bulk problem.
In the literature on coastal waves, a number of impressive analytical results exists in one dimensional configuration, assuming a straight coastline and bottom topography variations perpendicular to the coast \cite{leblond1981waves,zeitlin2018geophysical}.  Exhibiting topological features in this context allows one to predict which features of those one-dimensional spectra are robust to coastal deformations or to fluctuations of topography along the coast direction. This methods has already been proven extremely useful in condensed matter, photonics, acoustics, or mechanics \cite{HasanKane}. Here we show how these concepts can be transferred to coastal dynamics. A major difficulty comes from the fact that fluids are continuous media, by contrast with usual condensed matter systems admitting a lattice structure. This fundamental difference prohibits a straightforward application of  bulk-boundary correspondence theorems that make use of this lattice structure.}
 
{\color{black} Recently, the existence of two unidirectional shallow water wave modes trapped along the equator has been related to a topological invariant through a bulk-\textit{interface} correspondence \cite{delplace2017topological}, that can now be understood as a manifestation of Atiyah-Singer index theorem  \cite{Faure}. More precisely,  two modes of the equatorial wave spectrum transit from one wave band to another when the zonal (Eastward direction)  wavenumber $k_x$ is varied. This \textit{spectral flow} of two modes has been related to a {\color{black}(monopole)} Chern number$\chern=2\,$ that characterizes bulk eigenmodes twisting around a band-crossing point in parameter space $(k_x,k_y,f)$, with $k_y$ the meridional (Northward direction)  wavenumber, and $f$ the Coriolis parameter, that is twice the projection of the planet's {angular velocity}  onto the local vertical axis \cite{delplace2017topological}. The index$\chern\,\,$ describes how bulk properties are changed when $f$ is varied {\color{black}and changes sign. It thus yields topological information about the \textit{interface} wave problem defined by a change of sign of $f$}. 
This equatorial interface Chern number cannot be used to discuss coastal problems, as the Coriolis parameter is held fixed in those problems: coastal Kelvin waves and equatorial Kelvin waves belong to two different classes of problems.}
 
{\color{black} In condensed matter, wave problems involving the bulk-boundary correspondence usually make use of another topological invariant, that we denote here as the \textit{bulk Chern number} $C$. This bulk invariant is defined independently on each side of the interface between different materials that own a spectral gap}. In the equatorial case, this would amount to compute a bulk Chern number for each Hemisphere.  In this framework, the interface Chern number$\chern\,$ of the equator would just be given by the difference of bulk Chern numbers in each Hemispheres, for each waveband. 
The existence of a bulk Chern number $C$ computed for a given value of $f$ would then be suited to address the coastal problem. It turns out that this strategy is doomed, owing to the continuous nature of fluids. Indeed, the bulk Chern number $C$ is well defined for the wave bands of a two-dimensional condensed matter systems admitting an underlying lattice structure that makes compact the 2D parameter space $(k_x,k_y)$ over which this topological index is computed. This structure is lost in continuous media such as fluids, so that $C$ is in general ill-defined for eigenmodes parameterized on the plane $(k_x,k_y)$. Formally, a regularization parameter can be introduced to fix this problem \cite{tauber2019bulk,souslov2019topological}, but its introduction cannot be justified from first principles in the geophysical context. Besides, the bulk-boundary correspondence must be carefully stated to be valid in that case, since the value of the bulk Chern number $C$ does not necessarily correspond to the number of states that fill the gap in continuous media with a boundary \cite{tauber2020anomalous,graf2020topology}.

To bypass this difficulty, we follow in this paper a different strategy by showing that coastal Kelvin waves can be apprehended  {\color{black}{as an interface problem at fixed $f$ but with a varying bottom topography}}. We demonstrate their topological origin through the computation of an interface {\color{black} monopole Chern number$\chern=1$\, associated to a two-band crossing  point for bulk waves in parameter space. We introduce for that purpose a new parameter: the relative local gradient of bottom topography denoted $\beta_t$, whose importance has recently been highlighted in other shallow water wave transport problems \cite{onuki}. A change of sign of this parameter allows us to interpret the coast as an interface rather than a hard boundary.} This makes possible the use of bulk-interface correspondence machinery in $(k_x,k_y,\beta_t)$ parameter space, at fixed $f$. The historical Kelvin wave for an hard-wall boundary condition is then recovered as a limiting case of our theory, and other classes of coastal waves can be predicted and classified using this method.

The paper is organized as follows. We recall in section 2 a useful solvable case of rotating shallow water waves with varying bottom topography, and introduce the concept of spectral flow. We propose in section 2 a new formulation of the model on a convenient vectorial form to study topological properties of bulk eigenmodes. We show in section 3 that these topological properties explain a variety of spectra obtained for different problems in coastal dynamics. \textcolor{black}{Details on the actual computation of the Chern number of two-band degeneracy points are recalled in an appendix.} 

\section{Shallow water model for coastal waves}

\subsection{Shallow water model with bottom topography}
\begin{figure}
    \centering
    \includegraphics[width=0.8\columnwidth]{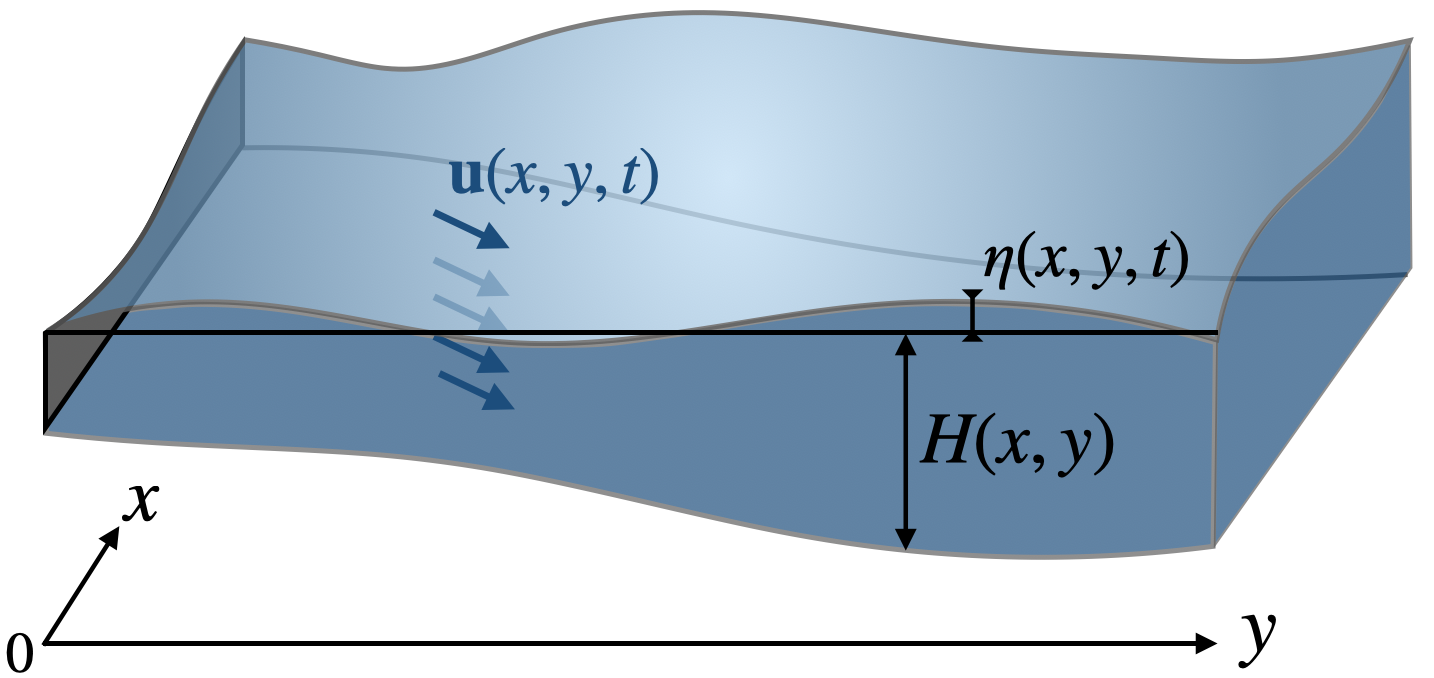}
    \caption{\textcolor{black}{Shallow fluid layer with a varying bottom topography $H(x,y)$. A perturbation around a state of rest induces a small elevation $\eta(x,y,t)$. The horizontal velocity  field $\mathbf{u}(x,y,t)$ (blue arrows) is uniform in the vertical direction.}
    \label{fig:SWtopog}}
\end{figure}

We consider a rotating shallow water model with varying bottom topography, that describes the dynamics of a thin fluid layer with homogeneous density \cite{VallisBook,zeitlin2018geophysical}, as sketched in figure \ref{fig:SWtopog}. 
In this approximation, the fluid is hydrostatic along the vertical direction, and the horizontal velocity  field $\mathbf{u}(x,y,t)=(u(x,y,t),v(x,y,t))$ is depth-independent. The layer thickness at a given point $(x,y)$ is $h=\eta+H(x,y)$ with  $\eta(x,y,t)$ the interface elevation around a state of rest. The bottom topography is thus encoded in the field $H(x,y)$. The shallow water dynamics is derived in the absence of dissipation effects using momentum conservation in the horizontal direction and mass conservation. The linearized dynamics around a state of rest is described by
\begin{align}
\partial_t \begin{pmatrix}
u \\
v \\
\eta
\end{pmatrix}= 
\begin{pmatrix}
0 & f & -g\partial_x \\
-f & 0& -g \partial_y  \\
-\partial_x (H \cdot)  & -\partial_y ( H \cdot )& 0     
\end{pmatrix} \begin{pmatrix}
u \\
v \\
\eta
\end{pmatrix}\label{eq:linear_shallow_water_init}
\end{align}
where $f$ is the Coriolis parameter and $g$ is the standard gravity constant at the surface of the planet. Time unit is chosen in the remaining of this paper such that $g=1$.

We review in the next subsection important previous results obtained in the sixties for particular bottom topography profiles. This allows us to introduce some standard terminology used in coastal dynamics, and to set the stage for a classification of coastal waves in a much more general framework, using tools from topology.

\subsection{A solvable case for coastal waves with varying bottom topography}

\begin{figure}
    \centering
    \includegraphics[width=0.8\columnwidth]{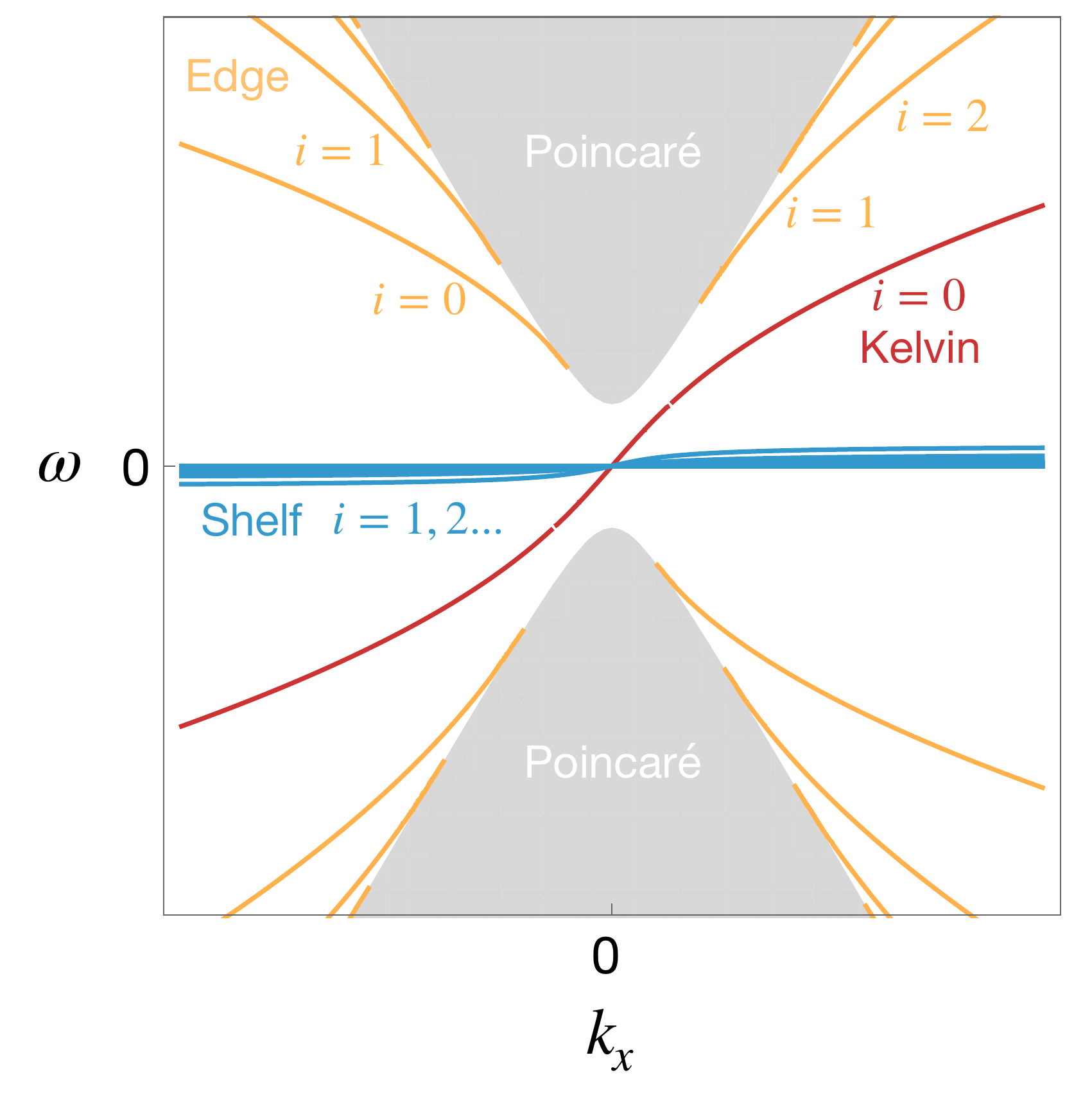}
    \caption{Dispersion relation of coastal rotating shallow water waves with an exponential ocean depth as in Eq. (\ref{eq:bal}), following \cite{ball1967edge}. This result is generic to other bottom topography profiles with a shore line $h(0)=0$\cite{huthnance1975trapped,leblond1981waves,zeitlin2018geophysical}. "Edge" refers to inertial-gravity trapped waves at the coast. "Shelf" refers to continental shelf waves trapped at the coast. The index $i$ gives the number of nodes in the wave amplitude. The grey region corresponds to continuous spectrum of delocalized inertia-gravity Poincar\'e  modes. The \textit{spectral flow} across  the frequency gap is one, and corresponds to the presence of the Kelvin wave.  
    \label{fig:Bal}}
\end{figure}

Depending on the topography profile at hand, a variety of wave spectra satisfying eq. (\ref{eq:linear_shallow_water_init}) have been previously described both analytically and numerically, together with some experimental observations of the corresponding waves, see e.g. \cite{huthnance1975trapped,hendershott1973ocean,leblond1981waves,zeitlin2018geophysical} and references therein. In particular,  Ball \cite{ball1967edge,reznik2011resonant} obtained analytical results in the case of a \textit{continental shelf} 
\begin{equation}
    H(y)=H_0\left(1-e^{-y/a}\right)
    \label{eq:bal}
\end{equation}
for $y>0$. The resulting spectrum is plotted in figure \ref{fig:Bal}. The results obtained in this particular case are generic to coastal problems where half a flat-bottom $f$ plane is connected to a shore line \cite{hendershott1973ocean,huthnance1975trapped,reznik2011resonant}.
Those configurations exhibit both discrete spectrum associated with trapped waves and continuous spectrum  associated with delocalized bulk waves. 

The non-zero frequency bulk waves are akin to those computed on an unbounded flat-bottom f-plane, and are therefore referred to as 
 \textit{Poincar\'e continuum}. This corresponds to two symmetric wavebands satisfying $\omega =\pm \sqrt{c^2 k_x^2+c^2 k_y^2 +f^2}$, with $(k_x,k_y)$ the wavenumber of the bulk waves and $c=\sqrt{gH_0}$ the phase speed of non-rotating shallow water gravity waves.  Those waves are also called \textit{inertia-gravity waves},  since they are surface waves influenced by rotation.

In the case of a continental shelf with the exponential profile \eqref{eq:bal}, Ball found  three additional classes of trapped waves. In the remaining of this paper, we shall follow  the  terminology used in \cite{huthnance1975trapped} for those waves:
\begin{itemize}
    \item \textit{Edge waves}: Inertia-gravity waves that are trapped  along the coast. Their frequency is always larger than the inertial one $f$. Edge waves are indexed by $i\in \mathbb{N}$ for $k_x<0$ and by  $i\in \mathbb{N}^*$ for $k_x>0$, with the index giving the number of nodes in the wave amplitude for the interface height variation. 
    \item \textit{Continental shelf waves}: Geostrophic modes trapped along the coast. Those shelf waves are indexed by  $i\in \mathbb{N}^*$ (whatever the value of $k_x)$, with $i$ giving the number of nodes in the wave amplitude for the interface height variation. Geostrophy means a balance between Coriolis force and pressure forces on the horizontal.
    \item \textit{Kelvin wave}: unidirectional wave whose frequency varies from $-\infty$ to $+\infty$ when increasing $k_x$ over the same range. Its dispersion relation thus  fills the frequency gap between the negative and positive Poincar\'e wavebands. The amplitude of the corresponding eigenmodes does not have any node. Following the convention for other edge waves, it is indexed by $i=0$. 
\end{itemize}

\subsection{Hard wall spectrum as a limiting case}

Textbooks in geophysical fluid dynamics usually present coastal waves by following the original work by Kelvin where the eponymous trapped wave is computed by considering a flat-bottom f-plane problem with a lateral wall  (see e.g. \cite{VallisBook}). This boundary imposes an impermeability constraint expressed as a condition of vanishing velocity across the wall. No other trapped wave than the Kelvin wave is found in this problem. In particular, there is neither inertia-gravity edge wave nor shelf  waves. 

Consistently, it was noticed by Ball that this original Kelvin's spectrum is recovered in the limit $a\rightarrow 0$ of an abrupt edge arbitrarily close to the vertical, for any given finite range of wavenumber $k_x$ in the direction along the wall. In this limit, the trapped shelf modes asymptote to the flat geostrophic continuum, the inertia-gravity edge modes have frequencies that tends to infinity, and the only remaining trapped mode with finite frequency is the Kelvin wave that connect the negative frequency Poincar\'e waveband to the positive frequency Poincar\'e waveband.

\subsection{Spectral flow}

Whatever the model for the coast, the dispersion relation of the Kelvin wave crosses the frequency gap between the geostrophic ($\omega=0$) and the Poincar\'e ($|\omega|>f$) continuum when varying $k_x$. This transition of a mode from a waveband to another when a parameter is varied is called a \textit{spectral flow}. 

In the following, we  argue that this spectral flow of Kelvin mode is a manifestation of  Atiyah-Singer theorem, which, in very loose terms, relates this  property to a topological index for bundles of complex  eigenmodes of a dual matrix problem that is much simpler to solve that the original linear wave operator \cite{nakahara2003geometry,Faure}. This simpler problem is refereed to  as the bulk problem in the following. 
\section{Bulk topographic shallow water waves}

\begin{figure*}
    \centering
     \includegraphics[width=0.8\textwidth]{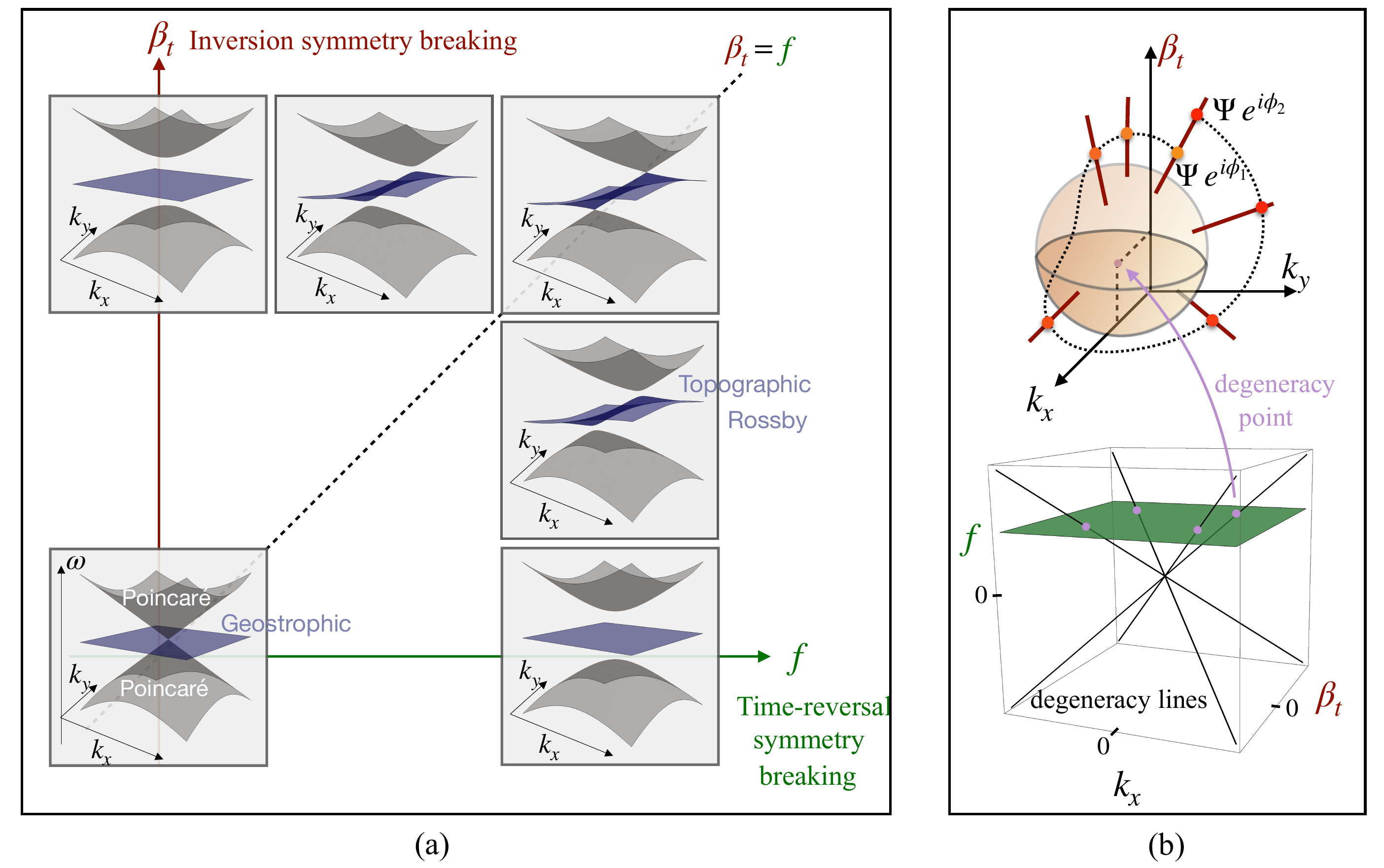}
\caption{  \label{fig:dispersion} (a) Dispersion relations of the bulk problem (\ref{eq:symbol}). The frequency gap between Poincar\'e wave bands and the topographic Rossby wave bands closes when $\beta_t=\pm f$, at $(k_x,k_y)=(\pm f/c,0)$.
(b) The upper panel sketches the vector bundle of shallow water complex eigenmodes  of the matrix (\ref{eq:symbol}), embedded in $(k_x,k_y,\beta_t)$ parameter space.  These eigenmodes are denoted $\Psi$. Once normalized, they are parametrized on a surface that encloses a double degeneracy point and have a phase freedom $\phi$ at each point of this surface. The impossibility to smoothly  define this phase over the sphere is a topological property of the vector bundle encoded through the interface Chern number$\chern$. (Lower panel) In the coastal case,  there are $4$ such degeneracy points $(k_x,k_y,\beta_t)=(\pm f/c,0,\pm f)$ to be considered, that result from the intersection of the double degeneracy lines of the bulk problem with the $f\neq0$ constant plane.}
  
\end{figure*}

\subsection{The bulk problem: symbol of the shallow water wave operator}

{\color{black} To define the bulk problem, we use a standard mapping between operators and their symbols provided by Weyl calculus \cite{Zworski}. Simply put, the symbol $H_{\text{bulk}}$ of an operator $\mathcal{H}_{op}$ for a multi-component wave problem such as in Eq. (\ref{eq:linear_shallow_water_init}) is a matrix that is obtained through a Wigner transform of the operator.  Similarly, it is possible to  define an operator from a symbol, using for instance Weyl calculus. See  \cite{Faure,onuki} for details and previous use of these transformation in the context of shallow water dynamics. In this framework, we get the following correspondence between derivatives and wavenumbers:
\begin{equation}
    \underbrace{c(y) \partial_y}_{\text{Operator}} \leftrightarrow \underbrace{i k_y c(y)  -\frac{1}{2}\partial_y c}_{\text{Symbol}} \label{eq:weyl}
\end{equation}
The study of symbols is at the heart of microlocal analysis that relates the solutions of of partial differential equation to ray tracing. Those tools were initially developed in the context of quantum physics and semi-classical analysis, see e.g. \cite{littlejohn1991geometric}. \textcolor{black}{This formalism  has recently  been proven fruitful in geophysical fluid dynamics; it has been used to prove the topological origin of equatorial waves \cite{Faure}, to explain the generic emergence of internal wave attractors in bounded stratified fluids \cite{colin}, and to describe wave transport properties in geophysical flows \cite{onuki}. This last application addressed specifically rotating shallow water waves with bottom topography. Our contribution is to unveil the role of topology in this problem.}

{\color{black} Before applying Weyl calculus, it is convenient to express the shallow water dynamics \eqref{eq:linear_shallow_water_init} as
\begin{align}
\partial_t \psi = -i\mathcal{H}_{op}\psi,\quad \psi\equiv (\sqrt{H} u, \sqrt{H} v,\sqrt{g} \eta)^t
\end{align}
\begin{align}
    \mathcal{H}_{op} \equiv i 
\begin{pmatrix}
0 & f & - c \partial_x \\
-f & 0& - c \partial_y  \\
-c  \partial_x  & -c  \partial_y - 2  \beta_{t}& 0     
\end{pmatrix} 
\label{eq:linear_shallow_water_change_op}
\end{align}}
where we have introduced the $y$-dependent wave phase speed and the relative gradient of bottom topography
\begin{equation}
      c\equiv\sqrt{gH}  , \quad \beta_{t}  \equiv\frac{1}{4} c \frac{\partial_y H}{H} \ . \label{eq:changevariable}
\end{equation}
{\color{black} To simplify the presentation, we restrict ourselves to a straight coast, with topography variations in the $y$ direction only. Topologically protected features obtained in this 1D case will be robust to smooth topography variations in the other direction \cite{HasanKane}.

Using Eq. (\ref{eq:weyl}) given by Weyl calculus, the symbol associated to the multicomponent linear wave operator (\ref{eq:linear_shallow_water_change_op}) is the Hermitian matrix
\begin{equation}
   H_{\text{bulk}}\equiv  \begin{pmatrix}
0 & if &  ck_x \\
-f & 0&  ck_y +i\beta_{t} \\
 ck_x  &  ck_y -i\beta_{t}& 0     \label{eq:symbol}
\end{pmatrix} 
\end{equation}}
{\color{black}Finding the eigenmodes $\hat{\psi}$ and eigenvalues $\omega$ of this matrix for a given set of parameters $k_x,k_y,\beta_t,c$ amounts to compute the (dual) bulk problem of rotating shallow water waves with varying bottom topography. This bulk problem is uniquely defined within the framework of Weyl calculus. The  corresponding eigenmodes are called bulk waves, as they can be interpreted as plane waves solutions of an unbounded wave problem where $\beta_{t}$ and $c$ are held constant, with $ik_x,i k_y$ replaced by $\partial_x,\partial_y$. 
As explained below, global properties of the shallow water wave spectrum with varying topography is encoded in the topological properties of the symbol (\ref{eq:symbol}).}

\subsection{Dispersion relation and discrete symmetries}

Before discussing shallow water wave spectrum  associated with a profile $\beta_t(y)$, and its relation with a coast, let us  first focus on the bulk problem, assuming $c$ constant, and $(k_x,k_y,\beta_t,f)$ taken as a set of free parameters.

The eigenmodes of the symbols correspond to three wavebands with frequencies satisfies
\begin{equation}
    \omega^3 -\omega \left( c^2 k_x^2+ c^2 k_y^2+f^2+\beta_t^2\right) + 2f\beta_t c k_x =0 \ .
    \label{eq:ploynome_dispersion}
\end{equation}

These wavebands are plotted in figure \ref{fig:dispersion} (a). In the flat bottom case ($\beta_t=0$) without rotation ($f=0$), the three wavebands consist in one zero-frequency
flat band  and two symmetric dispersionless modes  gravity wave modes of frequency $\pm\omega$ and of phase speed $c$. Those modes touch each other at $(k_x,k_y)=(0,0)$. In the unbounded $f$-plane case ($|f|>0$), with a flat bottom ($\beta_t=0$), a spectral gap of amplitude $|f|$ -- the inertial period -- separates the flat band from the gravity wave modes. In that case, the flat band modes are called geostrophic modes. The gravity wave modes influenced by rotations are called Poincar\'e or inertia-gravity wave modes.  In the $f$-plane situation ($f\neq 0$) with a  gradient of bottom topography ($\beta_t\neq 0$), the flat band acquires some dispersion.  Those low frequency waves bear strong similarities with  planetary Rossby waves encountered on a flat bottom ocean with a varying  Coriolis parameter.  \textcolor{black}{We therefore identify these modes as topographic Rossby waves \cite{leblond1981waves,zeitlin2018geophysical}.}

At the critical  value $|\beta_t|=|f|$, the topographic Rossby band touches  the Poincar\'e band, leading to a two-fold degeneracy point around which the dispersion relation is linear. This is the key observation that will allow us a topological analysis of the coastal Kelvin modes in the next section.  \textcolor{black}{The degeneracy points are systematically obtained by vanishing the discriminant of Eq (\ref{eq:ploynome_dispersion}), which leads to the conditions $k_y=0$ and $c^2 k_x^2=f^2=\beta_t^2$. This defines the subspace of two-fold degenerate eigenstates as lines in $(k_x,k_y=0,\beta_t,f)$ parameter space, as shown in figure $\ref{fig:dispersion}$ (c). These lines intersect each others at the origin $(k_y,k_x,f,\beta_t)=(0,0,0,0)$. This intersection corresponds to the three-fold waveband crossing point visible in the dispersion relations of figure \ref{fig:dispersion} (a). \cite{delplace2017topological}Here, by introducing a new bottom topography parameter, we describe the consequences of the degeneracy lines associated with two-fold degeneracy points, and discuss applications to coastal waves.}


The figure \ref{fig:dispersion} (a) reveals a striking parallel between the rotating shallow water model with a topography gradient and the celebrated Haldane model for Chern insulators \cite{haldane1988model}. This toy model, that turned out to be a building block of various topological materials, is based on a graphene lattice model for electrons. While the dispersion relation of graphene is known to show band crossings, called Dirac points, the Haldane model introduces additional terms that break either time-reversal or \textcolor{black}{mirror} symmetry. These symmetries breaking lead to two different gap opening mechanisms associated to two distinct topological phases. 
Similarly, here,  the Coriolis parameter $f$ breaks time-reversal symmetry and the topography gradient parameter $\beta_t$ breaks \textcolor{black}{mirror} symmetry in the $y$ direction. 
An important difference with the Haldane model 
is that the $(f,\beta_t)$ diagram in figure \ref{fig:dispersion} (a) is not a topological phase diagram as one can not assign a  bulk Chern number $C$ at each point.  In contrast, the topological aspect of our continuous model is expressed with a monopole Chern number$\chern$\, obtained by considering a variation of either $f$ or $\beta_t$ around a band crossing point, that occurs at $f=\beta_t$ (dashed line in figure \ref{fig:dispersion} (a)). 
In particular, for the coastal Kelvin wave we focus on, one needs to consider  a variation of $\beta_t$ at fixed $f$. Therefore, coastal Kelvin waves result from a concomitant breaking of both time-reversal and \textcolor{black}{mirror} symmetry. 

\subsection{Topology of eigenmodes around degeneracy points}

\textcolor{black}{The rotating shallow water model with a gradient of topography possesses topological properties associated to the aforementioned two-fold degeneracy points.  Each of these crossing points, when considered as isolated points in a three dimensional parameter space, are associated with a set of  monopole Chern numbers$\chern_n\, \in\mathbb{Z}$, where $n$ is the wave band index. In the coastal case we focus on, $f$ is fixed. The three dimensional parameter space to be considered is $(k_x,k_y,\beta_t)$, as depicted in figure \ref{fig:dispersion} (b).  It will convenient to use the more abstract notation $\boldsymbol{\lambda}=(\lambda_1,\lambda_2,\lambda_3)$  for this parameter space in the following paragraph.} 

 \textcolor{black}{The Chern numbers$\chern_n\, \in\mathbb{Z}$  describe the obstructions to smoothly define the arbitrary global phase of the normalized eigenstates $\Psi_n$ of  the bulk problem \eqref{eq:symbol} parameterized on a closed surface $\Sigma$ that encloses the degeneracy point in  parameter space $\boldsymbol{\lambda}$. In figure \ref{fig:dispersion} (b),  this surface is depicted as a sphere centered on one of the degeneracy point.}   For a given waveband $n$, the first Chern number is a global property of the eigenmode bundle defined by the closed base space $\Sigma$ and the set of eigenmodes $\Psi_n$ defined up to a phase on this base space  \cite{nakahara2003geometry}. It is an integer that somehow describes how  twisted the eigenmodes of a given waveband are on the close surface $\Sigma$. This numbers can be computed through a generalization of Gauss-Bonnet formula as 
\begin{equation}
    \chern_n=\frac{1}{2\pi} \int_{{\Sigma}} \mathbf{F}^{(n)}\cdot  \mathrm{d} \boldsymbol{\Sigma} \label{eq:chern}
\end{equation}
where $\mathbf{F}^{(n)}=(F^{(n)}_{\lambda_2,\lambda_3},F^{(n)}_{\lambda_3,\lambda_1},F^{(n)}_{\lambda_1,\lambda_2})$ is the Berry curvature, defined as 
\begin{equation}
    F^{(n)}_{\lambda_p,\lambda_q}= i \left(
    \frac{\partial \Psi_n^\dagger}{\partial {\lambda_p}} \cdot   \frac{\partial \Psi_n}{\partial {\lambda_q}} -\frac{\partial \Psi_n^\dagger}{\partial {\lambda_q}} \cdot  \frac{\partial \Psi_n}{\partial {\lambda_p}}\right) \label{eq:berry_curvature}
\end{equation} 
with the standard inner scalar product $\Psi_n^\dagger \cdot  \Phi_n=\sum_{j=1} \Psi_{nj}^* \Phi_{nj}$ where $\Psi_{nj}$ is the $j^{th}$ component of $\Psi_n$.}

\textcolor{black}{In fluid context, this formula was previously used to compute non-trivial  Chern numbers of shallow water eigenmode bundles enclosing the triple degeneracy point at the origin of $(k_x,k_y,f)$-space \cite{delplace2017topological}.  It is  tempting to interchange the roles of $f$ and $\beta_t$, by considering instead the monopole Chern numbers associated to  the triple degeneracy point at the origin of $(k_x,k_y,\beta_t)$-space. In that case, we find that the Chern numbers of the three wavebands vanish. The topography parameter alone opens a gap but does not induce non-trivial topology on eigenmode bundles. 
To find non-trivial topological properties, we need to consider $f\ne0$. This is the reason why we focus here on the Chern number of two-fold degeneracy points in $(k_x,k_y,\beta_t)$-parameter space.  Their analytical computation through Eq. (\ref{eq:chern}) requires in principle the expression of the eigenmodes of \eqref{eq:symbol}, which can be quite involved for a three-band problem. This difficulty is  bypassed here by focusing on the absolute value of$\chern$\, only. Since we deal with two-band crossing points around which  the dispersion relation is linear, we get $|\chern $ \ $|=1$ (see Appendix). We explain below how this number is related to the global shape of shallow water wave spectra with spatially varying bottom topography. }

\section{Spectral flow along coasts, abyss and escarpments }

\begin{figure*}
    \centering
    \includegraphics[width=0.8\textwidth]{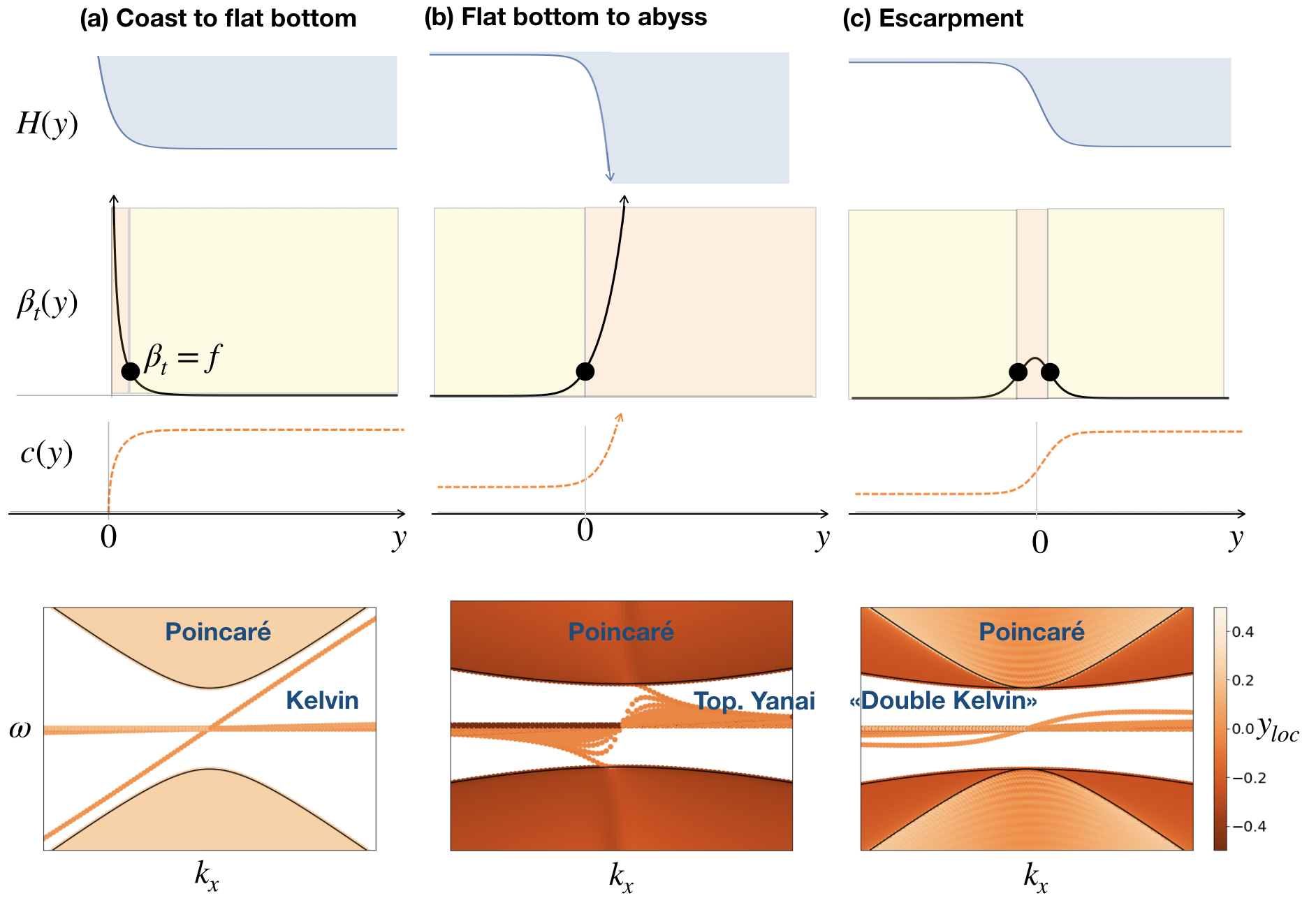}
    \caption{\label{fig:coastal_interfaces}  Three topologically distinct interface problems. (a) the case of a coast, where $\beta_t(y)$ crosses a degeneracy point $y_c>0$ from above, and $c$ vanishes at $y=0$. This is related to  spectral flow $+1$ due to the presence of the Kelvin wave in the spectrum below. (b) The case of a flat bottom f-plane opened to an "abyss" region where bottom topography diverges with $y\rightarrow +\infty$. The same degeneracy point as in case (a) is crossed by the profile $\beta_t(y)$, but from below. This is related to spectral flow $-1$ due to the presence of a topographic Yanai waves. (c) Case of an escarpment separating two flat bottom half f-planes with different depth. This is a topologically trivial case that can be interpreted as the concatenation of case (b) and (a). There is a so-called double Kelvin mode in the frequency gap; this mode is not associated with a spectral flow.}
\end{figure*}

We now argue that the topological properties of bulk degeneracy points can be used to explain the global shape of Ball's spectrum for coastal waves shown in figure \ref{fig:Bal}, and more generally to predict spectra for any kind of shallow water problems with varying bottom topography. We start by recalling several important mathematical results connecting the monopole Chern number$\chern$ \ to spectral flow in dual interface problems. 

\subsection{From degeneracy points to spectral flow \label{sub:spectral_flow}}

{\color{black}We now consider solutions of the wave equation \eqref{eq:linear_shallow_water_init} in a case where the topography $H(y)$ varies spatially. Consequently, both the parameters $\beta_t(y)$ and $c(y)$ vary with $y$.} When  $\beta_t(y)$ is a monotonic function that goes through a critical\footnote{\textcolor{black}{Note that this critical latitude should not be confused with critical layers that occur when waves propagate in the presence of a mean flow.}} latitude $y_c$ such that a degeneracy point exists for bulk waves in  ($k_x,k_y$,$\beta_t(y_c)$)-plane, one get an \textit{interface  problem} at $y_c$..

The existence of degeneracy points between bands with non trivial topological properties for the bulk eigenmodes in parameter space $(k_x,k_y,\beta_t)$ manifests itself in the  interface wave problem as a  spectral flow: some of the modes transit from one wave band to another when $k_x$ is varied \cite{Faure}. 
The number of modes that transit from one band $\omega_-$ to another one $\omega_+$ close to the degeneracy point, is (algebraically) equal to the monopole Chern number$\chern_+=-\chern_-$ associated with this degeneracy point. 
This method has been proven useful to interpret 
 molecular spectra \cite{faure2000topological}, to show the topological origin of equatorial \cite{delplace2017topological} and  Lamb-like waves in compressible stratified fluids \cite{perrot2019topological},  plasma \cite{parker2020topological,parker2020nontrivialb}, active matter flows \cite{shankar2017topological,green2020topological},  and to predict new electromagnetic modes in gyrotropic media \cite{marciani20}.

The state that transits from one band to another is trapped close to the critical value $y_c$ where the value of $\beta_t$ reaches the degeneracy point. The trapping length scale is generally given by an intrinsic length of the problem that plays the role of $\hbar$ in semi-classical analysis. In the present case, this trapping length scale is the Rossby radius of deformation $c/f$. 
When the profile $\beta_t(y)$ is associated with several degeneracy points, one can interpret qualitatively the global shape of the spectrum by considering these degeneracy points independently from each others, provided that the trapping  length scale $c/f$ is much smaller than the distance between two critical latitude $y_{c,i}$ and $y_{c,i+1}$ associated with different degeneracy points between the two bands.

For a given degeneracy point, the sign of the spectral flow changes with the sign of $\mathrm{d}  \beta_t /\mathrm{d}  y $. In other words, a state that transits from the lower band to the upper band when $\beta_t$ is an increasing function of $y$ would transit from the upper band to the lower band if $\beta_t(y)$ was a decreasing function of $y$ that crosses the same degeneracy point. When a given degeneracy point is crossed several time by the profile $\beta_t(y)$ over a distance smaller than the trapping length scale,  {two cases need to be considered, depending on the number of crossing points.  If this number is even, then there is no spectral flow.  If this number is odd, then the direction of the spectral flow is given by the one deduced from the first crossing point.}

Thus, topology provides a toolbox that makes possible a classification of coastal waves, when properly recast as an interface problem. In the following, we use this toolbox to interpret three classes of coastal spectra solved numerically using Dedalus software \cite{burns2020dedalus}. The results are summarized in figure \ref{fig:coastal_interfaces}.

\subsection{Topological origin of the coastal Kelvin wave}

The profile $\beta_t(y)$ of the relative topography gradient parameter associated with the exponential Ball's bottom topography profile $H(y)$ of Eq. (\ref{eq:bal}) is shown in figure \ref{fig:coastal_interfaces} (a). The key point is that $\beta_t(y)$ is decreasing from $y=0$ to $y\rightarrow +\infty$, and  goes through the degeneracy point $\beta_t=f$ with $\beta_t'<0$. We  have thus recast the coastal problem into an interface problem. 

As expected from our analysis of the bulk problem, and from general results on the correspondence between such bulk indices and spectral flow for the dual interface problems, we observe that the spectrum of figure \ref{fig:coastal_interfaces} (a) exhibits one state that transits from the lower frequency Poincaré wave band to the upper one. This shows the topological origin of the coastal Kelvin wave. This topological property distinguishes the coastal Kelvin wave from the other trapped boundary waves, namely edge and shelf waves, as it guarantees its existence against  continuous deformations of the topography profile and constrains its dispersion relation to fill the frequency gap.

We stress that the interface is not the coast itself (at $y=0$); it is  rather the critical point $y_c$ where $\beta_t(y_c)=f$. In the limit $a\rightarrow 0$, this critical point becomes closer to the coast: $y_c\rightarrow 0$.

Depending on the detailed shape of $H(y)$, they could be actually an odd number of degeneracy points, but the difference between degeneracy points associated with $\beta'_t<0$ and degeneracy points associated with  $\beta'_t>0$ will be always one.  We also notice  that the intrinsic length scale vanishes close to the coast as  $c(y)/f\sim y^{1/2}$. Qualitatively, this guarantees that the actual  shore line is screened from the interior dynamics.

\subsection{Topographic Yanai waves at an interface with deep waters}

In the previous case, we found that one mode is gained by the upper band when $k_x$ is increased, and related this spectral flow to a profile $\beta_t(y)$ crossing the degeneracy point from above. We expect an opposite spectral flow when the profile $\beta_t(y)$ crossed the same degeneracy point from below. That is to say, we expect a net loss of one mode in the upper band when $k_x$ is increased. This is the situation depicted in figure \ref{fig:coastal_interfaces} (b), with the topography profile  
$H(y)=H_0\left(1+3 e^{-\frac{y_{a}-y}{0.1}}\right)$. 
This profile  is representative of cases where half a flat-bottom $f$-plane opens to infinitely deep water (the abyss). 

As shown in the spectrum displayed in figure \ref{fig:coastal_interfaces} (b), there is indeed a mode that transits from the upper Poincar\'e band to the geostrophic wave band when $k_x$ is increased. 
This topological mode is an inertial mode at $k_x=0$, where $\omega=f$. Its branch bears strong similarities with the equatorial Yanai mode, \textcolor{black}{as is connects the topographic Rossby wave band to the inertia-gravity waveband}.  We therefore propose to call it a \textit{topographic Yanai mode}; the existence of such modes was actually described in previous work by Iga in a  classification of shallow water spectra depending on the boundary conditions in channel geometries \cite{iga1995transition}. 
Our study now shows that such a mode have a topological origin, just as the coastal  Kelvin wave. As far as topology is concerned, the only difference between those two modes is their opposite group velocity, that can be related to the different sign of $\beta_t'$ at the critical point where $\beta_t(y_c)=f$.

\textcolor{black}{The topography profile $H(y)$ related to the topographic Yanai case is peculiar, as it diverges with $y$. While a number of interesting and useful results are often derived in a shallow water context with diverging bottom topography profiles, one should keep in mind that such profile are not consistent with the hypothesis underlying the derivation of shallow water equations from more comprehensive 3D Euler dynamics. It will be interesting to ask wether topographic Yanai surface waves exist in the context of incompressible 3D Euler flows with gravity. Within the shallow water  framework, the connection from a shallow coastal area to a deeper abyss-like area can only be consistent with the hypothesis required for the model derivation if the abyss depth $H(y)$ tends to a constant value at large $y$. In other words, the profile $\beta_t(y)$ must be decreasing towards the origin at large $y$. This leads to the escarpment case discussed below.}

\subsection{Double Kelvin waves over escarpment are topologically trivial}

An \textit{escarpment} is a variation of topography separating two oceanic basins with different depths, as illustrated in figure \ref{fig:coastal_interfaces} (c). We argue below that escarpments result from the concatenation of the two topological cases described above. Since the singularities cancel each other, this leads to a configuration that is topologically trivial. This leads however to a non-trivial reinterpretation of the spectra associated with escarpments.

In the case of an escarpment $H(y)$ separating a shallow oceanic basin with depth $h_1$ to a deep oceanic basin with depth $h_2$, a peculiar trapped mode with frequency lower than the inertial frequency $f$, and without node in their amplitude in the $y$ direction, was found in the sixties by Longuet-Higgins \cite{longuet1968double,longuet1968trapping}. This mode was dubbed \textit{double Kelvin wave}, because its amplitude is decaying on both side of the escarpment. 
Those waves do not fill the gap between geostrophic mode and Poincar\'e modes, as reported in figure \ref{fig:coastal_interfaces} (c).  Instead, their dispersion relation is similar to Rossby waves, with a group velocity changing sign at high wavenumbers.  In particular, these trapped modes are not unidirectional.

From a topology and spectral flow perspective, the absence of modes transiting from one band to another is consistent with the profile $\beta_t(y)$ that  either crosses twice the same degeneracy point or does cross any degeneracy point at all when the escarpment is not sufficiently steep.
The case of a double crossing of the same degeneracy point with opposite sgn($\beta'_t$) is topologically equivalent to the one without degeneracy points. \textcolor{black}{The double Kelvin wave is topologically trivial. Our analysis shows that this peculiar mode can now be understood as a hybrid mode between a coastal Kelvin wave related to the first degeneracy point and a coastal topographic Yanai wave related to the second degeneracy point. The singularities associated with those two degeneracy points cancel each other. In other words, the double Kelvin mode of case (c) in figure \ref{fig:coastal_interfaces} is a footprint of the concatenation of $\beta_t$-profiles described in cases (a) and (b). As such, those hybrid modes could rather be dubbed \textit{mixed topographic Yanai-Kelvin} waves. By this way, topology gives a complementary point of view to \cite{iga1995transition} on this problem.}

\section{Conclusion}

Topology makes possible a new classification of shallow water waves with varying topography.
This unifies several results on coastal waves reported in previous work, e.g. \cite{leblond1981waves,zeitlin2018geophysical,iga1995transition}. Topology guarantees that the spectral flow reported in the case of a straight coastline and one-dimensional topography variations in the direction perpendicular to the coast are also robust to changes in coastline shape or to perturbation of the topography profile in the direction along the coastline, as observed in experiments \cite{ren2021robust}.  

\textcolor{black}{It is worth stressing similarities and differences between the use of topology in this study and in condensed matter physics. Topology is routinely used to characterize  trapped boundary modes in condensed matter or related systems admitting an underlying lattice structure, via the celebrated bulk-boundary correspondence. Strictly speaking, this correspondence cannot be applied in continuous media such as fluid systems in general, unless some regularization terms are present in the physical model, as it might be the case e.g. in certain active fluids \cite{souslov2019topological,tauber2019bulk,tauber2020anomalous}. This procedure allows a rigorous investigation of the bulk-boundary correspondence, accounting for various boundary conditions beyond the impermeability one, when the dispersion relation is gapped due to the introduction of a fixed parameter that breaks a symmetry, such as $f$  \cite{tauber2020anomalous,graf2020topology}}.

\textcolor{black}{
Although the introduction of the regularization term could be meaningful in active matter systems, it is rather artificial in the geophysical realm. To avoid the use of such a mathematical trick, we found a quite natural geophysical way,  in this study, to define a Chern number in the $f$-plane, by considering a parameter $\beta_t$ in the equation that introduces new degeneracy points in the dispersion relation in a 3D parameter space. The introduction of this parameter does not regularize the eigenmode vector field at infinity. Instead, it can be seen as a regularization parameter of the abrupt wall considered in \cite{tauber2019bulk,tauber2020anomalous}. Concretely, this procedure turns the coastal (boundary) problem into an interface problem, where monopole Chern numbers appear. The key difference with the usual bulk-boundary correspondence being that $\beta_t$ is a varying parameter that changes sign, and not simply a fixed term that opens a gap. Those Chern monopoles can naturally be used in fluids problems or in other continuous media to make predictions about interface geometries (i.e. partial differential equations  with spatially varying parameter, without a boundary), by means of the Atiyah-Singer index theorem. This contrasts with other demonstrations of the bulk-boundary correspondence in the condensed matter context that involve an actual (sharp) boundary, and that do not refer to the Atiyah-Singer index theorem \cite{hatsugai1993chern,grafbulkedge2018,ProdanEmilSchultz}.}

{\color{black} Using this classification, it is now possible to discover new  classes of ``topographic-equatorial'' waves, when both parameters $f$  and $\beta_t$ vary in the meridional direction. In particular, since the critical latitudes associated with degeneracy points are shifted by the presence of a relative topographic gradient $\beta_t\ne 0$, we predict that the location of trapped modes will be also shifted. This is a new example of a dynamical equator that differs from the usual equator \cite{boyd2018dynamics}.} It will be also interesting to investigate in future works the dynamics of relatively deep planetary waves close to the tangent equatorial cylinder in Giant planets such as Jupiter \cite{heimpel2005simulation}. In fact, this case bear strong similarities with the topography profile "f-plane to abyss" in figure \ref{fig:coastal_interfaces} (b), with possible experimental realisation where the parameter $\beta_t/f$ has already been shown to play a central role \cite{cabanes2017laboratory}.  

The present work dealt with linear waves around a state of rest; this sets the stage for more comprehensive studies addressing the role of topology in the presence of nonlinearities, with a possible two way coupling between coastal boundary waves and interior dynamics \cite{deremble2017coupled,venaille2020quasi}.

From a fundamental perspective, the bottom line of this study is an interpretation of the Kelvin's original boundary problem with an impermeable wall as a limiting case of an interface problem with a varying bottom topography. This unveils the topological origin of the unidirectional trapped mode  computed by Kelvin in his seminal 1880 paper on tides, as resulting from an interplay between time-reversal and mirror symmetry breaking.\\

\appendix*
\section{Topology of eigenmode bundles enclosing two-band degeneracy points.}

{\color{black} We derive here a classical result on the topology of eigenmodes bundles enclosing two-band degeneracy points. More details are presented in standard reviews and textbooks on the physics of topological waves. We consider here two-band crossing of an Hermitian problem parameterized by a vector $\boldsymbol{\lambda}=(\lambda_1,\lambda_2,\lambda_3)$ that vanishes at the band-crossing point. In our case $\boldsymbol{\lambda}=(k_x,k_y,\beta_{t,y})-(\pm f, 0,\pm f)$. After a projection onto the two bands crossing each other, the two-band crossing problem is described by a reduced $2\times2$ matrix, which can always be written as
\begin{equation}
    H_r(\boldsymbol{\lambda})= \begin{pmatrix} h_3(\boldsymbol{\lambda})  + \omega_0& h_1(\boldsymbol{\lambda}) -ih_2 (\boldsymbol{\lambda}) \\
    h_1(\boldsymbol{\lambda}) +ih_2 (\boldsymbol{\lambda}) & -h_3(\boldsymbol{\lambda}) +\omega_0
    \end{pmatrix}\label{eq:eigenproblem}
\end{equation}
where $\boldsymbol{h}(\boldsymbol{\lambda})=(h_1(\boldsymbol{\lambda}), h_2(\boldsymbol{\lambda}), h_3(\boldsymbol{\lambda})) \in \mathbb{R}^3$. The eigenvalues of the matrix  $H_r$ are given by $\omega_\pm =\omega_0 \pm\sqrt{h_1^2(\boldsymbol{\lambda}) + h_2^2(\boldsymbol{\lambda}) + h_3^2(\boldsymbol{\lambda})}$ and cross at $\omega_\pm=\omega_0$ when $\mathbf{h}=0$.

Let us for a moment forget the dependence on $\boldsymbol{\lambda}$, and consider $\mathbf{h}=(h_1,h_2,h_3)$ as a parameter. Using Eq. (\ref{eq:chern}-\ref{eq:berry_curvature}), it is a classical exercise to show that the Chern number of the two eigenmode bundles parameterized over a closed surface $\Sigma_{\mathbf{h}}$ enclosing the degeneracy point $\mathbf{h}=0$ are 
\begin{equation}
    \mathcal{C}^h_\pm= \frac{1}{2\pi} \int_{\Sigma_{\mathbf{h}}} \mathbf{F}^{(\pm)}(\mathbf{h}) \cdot \mathrm{d} \boldsymbol{\Sigma}_{\mathbf{h}}=\mp 1 ,
\end{equation}
where $\mathbf{F}^{(\pm)}$ is the Berry curvature associated with eigenmodes of Eq. (\ref{eq:eigenproblem}) denoted $\Psi_{\pm}(\mathbf{h})$. The computation is explained in detail in \cite{fruchart2013introduction,Faure,bernevig}. One can actually visualize the singularity associated with this non-zero Chern number by a direct inspection of the eigenmode expression, using polar coordinates $(h,\theta,\phi)$ in parameter space with  $h_1=h\sin\theta \cos \phi$, $h_2=h\sin\theta \sin \phi$, $h_3=h \cos \theta$:
\begin{equation}
\Psi_+=\begin{pmatrix}
e^{-i \phi}\cos\frac{\theta}{2} \\ \sin \frac{\theta}{2}
\end{pmatrix}, \quad \Psi_-=\begin{pmatrix}
-\sin\frac{\theta}{2} \\e^{i \phi} \cos \frac{\theta}{2}
\end{pmatrix} .
\end{equation}
Those normalized eigenmodes are defined up to an arbitrary phase factor $e^{i\alpha_{\pm}}$.
With our phase choices, the singularity in $\Psi_\pm(\mathbf{h})$ occurs at $\theta=\pi$, as $\mathbf{h} $ is left invariant by changes in $\phi$, while $\Psi_\pm$ varies with $\phi$. One can use the phase freedom to remove the singularity from the location where $\theta=\pi$.  However, this phase change would only shift the singularity elsewhere on the eigenmode bundle defined on any surface enclosing the origin $h=0$. We also see that at the singular point $\theta=\pi$, the phase factor varying with $\phi$ is opposite for  $\Psi_-$ and for $\Psi_+$. The Chern number $\mathcal{C}^h_{\pm}$ quantifies these singularities. %

Now, we want to compute the Chern number for the eigenmode bundles  defined on a surface denoted $\Sigma_{\boldsymbol{\lambda}}$ in $\boldsymbol{\lambda}$-parameter space  rather than  in $\mathbf{h}$-space. We assume that the surface $\Sigma_{\boldsymbol{\lambda}}$ encloses the band crossing point $\boldsymbol{\lambda}=0$. We introduce the degree $\deg h$ that counts how many times the application  $h:\ \boldsymbol{\lambda}\in \Sigma_{\boldsymbol{\lambda}} \rightarrow \boldsymbol{h}/|\boldsymbol{h}| \in \Sigma_{\boldsymbol{h}}$ wraps the unit sphere in $\mathbf{h}$-parameter space when $\boldsymbol{\lambda}$ is varied over $\Sigma_{\boldsymbol{\lambda}}$. A direct computation of the Chern number through the integral of Berry curvature expressed either in $\boldsymbol{\lambda}$-parameter space or $\mathbf{h}$-parameter space yields to 
\begin{eqnarray}
    \chern_{\pm}&=&\frac{1}{2\pi} \int_{\Sigma_{\boldsymbol{\lambda}}} \mathbf{F}^{(\pm)}(\boldsymbol{\lambda}) \mathrm{d}{\boldsymbol{\Sigma}}_{\boldsymbol{\lambda}} \\
    &=& \frac{1}{2\pi} \int_{h(\Sigma_{\boldsymbol{\lambda}})}  \mathbf{F}^{(\pm)}(\boldsymbol{h}) \mathrm{d} \boldsymbol{\Sigma}_{\boldsymbol{h}} =  \left(\deg h\right) \mathcal{C}^h_\pm 
\end{eqnarray}
The last equality is obtained by noting that integrating the Berry curvature over a closed surface $\Sigma_{\mathbf{h}}$ enclosing $\mathbf{h}=0$ yields the same results as integration over the unit sphere in $\mathbf{h}$-space.

A method to find  $\deg h$ is to consider an arbitrary vector $\mathbf{h}_0 $, to find all the vectors $\boldsymbol{\lambda}_0$  such that $\boldsymbol{\lambda}_0=  \mathbf{h}^{-1}(\mathbf{h_0})$, and to compute
\begin{align}\label{eq:degree}
	\deg h = \sum_{\boldsymbol{\lambda}_0\in \boldsymbol{h}^{-1}(\mathbf{h}_0)} \text{sgn}\left[\det\left(\dfrac{\partial h_j}{\partial \lambda_{i}}\right)\bigg|_{\boldsymbol{\lambda}_0}\right] ,
	\end{align}
see e.g. \cite{DubrovinBook} for more details on the degree of an application.  Generically, the functions  $\{{h}_j\}$ depends linearly on the parameters $\{\lambda_i\}$ when $\boldsymbol{\lambda}\rightarrow 0$, i.e. when $\Sigma_{\boldsymbol{\lambda}}$ is chosen sufficiently close to the degeneracy point. In that case, there is a unique vector $\boldsymbol{\lambda}_0=h^{-1}$ to be taken into account in the sum, so that the possible values of the degree are restricted to $\pm 1$. In other words, close to the degeneracy point there is a non-singular linear transformation from $\boldsymbol{\lambda}$ to $\mathbf{h}$, so that $\mathbf{h}/|\mathbf{h}|$ wraps one time the sphere $S^2$ when $\boldsymbol{\lambda}$ wraps the surface $\Sigma_{\boldsymbol{\lambda}}$, and the degree accounts for a possible change orientation induced by the linear transformation. Consequently, the Chern numbers of the eigenmode bundles enclosing the degeneracy point are the same up to a sign in $\boldsymbol{\lambda}$-parameter space and $\mathbf{h}$-parameter space.}

\begin{acknowledgments}
We warmly thank Frederic Faure, Yohei Onuki, Nicolas Perez and  Clement Tauber for their useful insights on this problem.
\end{acknowledgments}

\bibliography{topol_topog}

\begin{thebibliography}{45}%
\makeatletter
\providecommand \@ifxundefined [1]{%
 \@ifx{#1\undefined}
}%
\providecommand \@ifnum [1]{%
 \ifnum #1\expandafter \@firstoftwo
 \else \expandafter \@secondoftwo
 \fi
}%
\providecommand \@ifx [1]{%
 \ifx #1\expandafter \@firstoftwo
 \else \expandafter \@secondoftwo
 \fi
}%
\providecommand \natexlab [1]{#1}%
\providecommand \enquote  [1]{``#1''}%
\providecommand \bibnamefont  [1]{#1}%
\providecommand \bibfnamefont [1]{#1}%
\providecommand \citenamefont [1]{#1}%
\providecommand \href@noop [0]{\@secondoftwo}%
\providecommand \href [0]{\begingroup \@sanitize@url \@href}%
\providecommand \@href[1]{\@@startlink{#1}\@@href}%
\providecommand \@@href[1]{\endgroup#1\@@endlink}%
\providecommand \@sanitize@url [0]{\catcode `\\12\catcode `\$12\catcode
  `\&12\catcode `\#12\catcode `\^12\catcode `\_12\catcode `\%12\relax}%
\providecommand \@@startlink[1]{}%
\providecommand \@@endlink[0]{}%
\providecommand \url  [0]{\begingroup\@sanitize@url \@url }%
\providecommand \@url [1]{\endgroup\@href {#1}{\urlprefix }}%
\providecommand \urlprefix  [0]{URL }%
\providecommand \Eprint [0]{\href }%
\providecommand \doibase [0]{https://doi.org/}%
\providecommand \selectlanguage [0]{\@gobble}%
\providecommand \bibinfo  [0]{\@secondoftwo}%
\providecommand \bibfield  [0]{\@secondoftwo}%
\providecommand \translation [1]{[#1]}%
\providecommand \BibitemOpen [0]{}%
\providecommand \bibitemStop [0]{}%
\providecommand \bibitemNoStop [0]{.\EOS\space}%
\providecommand \EOS [0]{\spacefactor3000\relax}%
\providecommand \BibitemShut  [1]{\csname bibitem#1\endcsname}%
\let\auto@bib@innerbib\@empty
\bibitem [{\citenamefont {Thomson}(1880)}]{kelvin}%
  \BibitemOpen
  \bibfield  {author} {\bibinfo {author} {\bibfnamefont {W.}~\bibnamefont
  {Thomson}},\ }\bibfield  {title} {\bibinfo {title} {On gravitational
  oscillations of rotating water},\ }\href@noop {} {\bibfield  {journal}
  {\bibinfo  {journal} {Proceedings of the Royal Society of Edinburgh}\
  }\textbf {\bibinfo {volume} {10}},\ \bibinfo {pages} {92} (\bibinfo {year}
  {1880})}\BibitemShut {NoStop}%
\bibitem [{\citenamefont {Hatsugai}(1993)}]{hatsugai1993chern}%
  \BibitemOpen
  \bibfield  {author} {\bibinfo {author} {\bibfnamefont {Y.}~\bibnamefont
  {Hatsugai}},\ }\bibfield  {title} {\bibinfo {title} {Chern number and edge
  states in the integer quantum hall effect},\ }\href@noop {} {\bibfield
  {journal} {\bibinfo  {journal} {Physical review letters}\ }\textbf {\bibinfo
  {volume} {71}},\ \bibinfo {pages} {3697} (\bibinfo {year}
  {1993})}\BibitemShut {NoStop}%
\bibitem [{\citenamefont {LeBlond}\ and\ \citenamefont
  {Mysak}(1981)}]{leblond1981waves}%
  \BibitemOpen
  \bibfield  {author} {\bibinfo {author} {\bibfnamefont {P.~H.}\ \bibnamefont
  {LeBlond}}\ and\ \bibinfo {author} {\bibfnamefont {L.~A.}\ \bibnamefont
  {Mysak}},\ }\href@noop {} {\emph {\bibinfo {title} {Waves in the Ocean}}}\
  (\bibinfo  {publisher} {Elsevier},\ \bibinfo {year} {1981})\BibitemShut
  {NoStop}%
\bibitem [{\citenamefont {Zeitlin}(2018)}]{zeitlin2018geophysical}%
  \BibitemOpen
  \bibfield  {author} {\bibinfo {author} {\bibfnamefont {V.}~\bibnamefont
  {Zeitlin}},\ }\href@noop {} {\emph {\bibinfo {title} {Geophysical fluid
  dynamics: understanding (almost) everything with rotating shallow water
  models}}}\ (\bibinfo  {publisher} {Oxford University Press},\ \bibinfo {year}
  {2018})\BibitemShut {NoStop}%
\bibitem [{\citenamefont {Hasan}\ and\ \citenamefont {Kane}(2010)}]{HasanKane}%
  \BibitemOpen
  \bibfield  {author} {\bibinfo {author} {\bibfnamefont {M.~Z.}\ \bibnamefont
  {Hasan}}\ and\ \bibinfo {author} {\bibfnamefont {C.~L.}\ \bibnamefont
  {Kane}},\ }\bibfield  {title} {\bibinfo {title} {Colloquium: Topological
  insulators},\ }\href {https://doi.org/10.1103/RevModPhys.82.3045} {\bibfield
  {journal} {\bibinfo  {journal} {Rev. Mod. Phys.}\ }\textbf {\bibinfo {volume}
  {82}},\ \bibinfo {pages} {3045} (\bibinfo {year} {2010})}\BibitemShut
  {NoStop}%
\bibitem [{\citenamefont {Delplace}\ \emph {et~al.}(2017)\citenamefont
  {Delplace}, \citenamefont {Marston},\ and\ \citenamefont
  {Venaille}}]{delplace2017topological}%
  \BibitemOpen
  \bibfield  {author} {\bibinfo {author} {\bibfnamefont {P.}~\bibnamefont
  {Delplace}}, \bibinfo {author} {\bibfnamefont {J.}~\bibnamefont {Marston}},\
  and\ \bibinfo {author} {\bibfnamefont {A.}~\bibnamefont {Venaille}},\
  }\bibfield  {title} {\bibinfo {title} {Topological origin of equatorial
  waves},\ }\href@noop {} {\bibfield  {journal} {\bibinfo  {journal} {Science}\
  }\textbf {\bibinfo {volume} {358}},\ \bibinfo {pages} {1075} (\bibinfo {year}
  {2017})}\BibitemShut {NoStop}%
\bibitem [{\citenamefont {Faure}(2019)}]{Faure}%
  \BibitemOpen
  \bibfield  {author} {\bibinfo {author} {\bibfnamefont {F.}~\bibnamefont
  {Faure}},\ }\bibfield  {title} {\bibinfo {title} {Manifestation of the
  topological index formula in quantum waves and geophysical waves},\
  }\href@noop {} {\bibfield  {journal} {\bibinfo  {journal} {arXiv}\ }\textbf
  {\bibinfo {volume} {1901.10592}} (\bibinfo {year} {2019})}\BibitemShut
  {NoStop}%
\bibitem [{\citenamefont {Tauber}\ \emph {et~al.}(2019)\citenamefont {Tauber},
  \citenamefont {Delplace},\ and\ \citenamefont {Venaille}}]{tauber2019bulk}%
  \BibitemOpen
  \bibfield  {author} {\bibinfo {author} {\bibfnamefont {C.}~\bibnamefont
  {Tauber}}, \bibinfo {author} {\bibfnamefont {P.}~\bibnamefont {Delplace}},\
  and\ \bibinfo {author} {\bibfnamefont {A.}~\bibnamefont {Venaille}},\
  }\bibfield  {title} {\bibinfo {title} {A bulk-interface correspondence for
  equatorial waves},\ }\href@noop {} {\bibfield  {journal} {\bibinfo  {journal}
  {Journal of Fluid Mechanics}\ }\textbf {\bibinfo {volume} {868}} (\bibinfo
  {year} {2019})}\BibitemShut {NoStop}%
\bibitem [{\citenamefont {Souslov}\ \emph {et~al.}(2019)\citenamefont
  {Souslov}, \citenamefont {Dasbiswas}, \citenamefont {Fruchart}, \citenamefont
  {Vaikuntanathan},\ and\ \citenamefont {Vitelli}}]{souslov2019topological}%
  \BibitemOpen
  \bibfield  {author} {\bibinfo {author} {\bibfnamefont {A.}~\bibnamefont
  {Souslov}}, \bibinfo {author} {\bibfnamefont {K.}~\bibnamefont {Dasbiswas}},
  \bibinfo {author} {\bibfnamefont {M.}~\bibnamefont {Fruchart}}, \bibinfo
  {author} {\bibfnamefont {S.}~\bibnamefont {Vaikuntanathan}},\ and\ \bibinfo
  {author} {\bibfnamefont {V.}~\bibnamefont {Vitelli}},\ }\bibfield  {title}
  {\bibinfo {title} {Topological waves in fluids with odd viscosity},\
  }\href@noop {} {\bibfield  {journal} {\bibinfo  {journal} {Physical {R}eview
  {L}etters}\ }\textbf {\bibinfo {volume} {122}},\ \bibinfo {pages} {128001}
  (\bibinfo {year} {2019})}\BibitemShut {NoStop}%
\bibitem [{\citenamefont {Tauber}\ \emph {et~al.}(2020)\citenamefont {Tauber},
  \citenamefont {Delplace},\ and\ \citenamefont
  {Venaille}}]{tauber2020anomalous}%
  \BibitemOpen
  \bibfield  {author} {\bibinfo {author} {\bibfnamefont {C.}~\bibnamefont
  {Tauber}}, \bibinfo {author} {\bibfnamefont {P.}~\bibnamefont {Delplace}},\
  and\ \bibinfo {author} {\bibfnamefont {A.}~\bibnamefont {Venaille}},\
  }\bibfield  {title} {\bibinfo {title} {Anomalous bulk-edge correspondence in
  continuous media},\ }\href@noop {} {\bibfield  {journal} {\bibinfo  {journal}
  {Physical Review Research}\ }\textbf {\bibinfo {volume} {2}},\ \bibinfo
  {pages} {013147} (\bibinfo {year} {2020})}\BibitemShut {NoStop}%
\bibitem [{\citenamefont {Graf}\ \emph {et~al.}(2020)\citenamefont {Graf},
  \citenamefont {Jud},\ and\ \citenamefont {Tauber}}]{graf2020topology}%
  \BibitemOpen
  \bibfield  {author} {\bibinfo {author} {\bibfnamefont {G.~M.}\ \bibnamefont
  {Graf}}, \bibinfo {author} {\bibfnamefont {H.}~\bibnamefont {Jud}},\ and\
  \bibinfo {author} {\bibfnamefont {C.}~\bibnamefont {Tauber}},\ }\bibfield
  {title} {\bibinfo {title} {Topology in shallow-water waves: a violation of
  bulk-edge correspondence},\ }\href@noop {} {\bibfield  {journal} {\bibinfo
  {journal} {arXiv}\ }\textbf {\bibinfo {volume} {2001.00439}} (\bibinfo {year}
  {2020})}\BibitemShut {NoStop}%
\bibitem [{\citenamefont {Onuki}(2020)}]{onuki}%
  \BibitemOpen
  \bibfield  {author} {\bibinfo {author} {\bibfnamefont {Y.}~\bibnamefont
  {Onuki}},\ }\bibfield  {title} {\bibinfo {title} {Quasi-local method of wave
  decomposition in a slowly varying medium},\ }\href@noop {} {\bibfield
  {journal} {\bibinfo  {journal} {Journal of Fluid Mechanics}\ }\textbf
  {\bibinfo {volume} {883}} (\bibinfo {year} {2020})}\BibitemShut {NoStop}%
\bibitem [{\citenamefont {Vallis}(2017)}]{VallisBook}%
  \BibitemOpen
  \bibfield  {author} {\bibinfo {author} {\bibfnamefont {G.~K.}\ \bibnamefont
  {Vallis}},\ }\href@noop {} {\emph {\bibinfo {title} {Atmospheric and Oceanic
  Fluid Dynamics~: Fundamentals and Large-Scale Circulation}}},\ \bibinfo
  {edition} {2nd}\ ed.\ (\bibinfo  {publisher} {Cambridge University Press,
  Cambridge, U. K.},\ \bibinfo {year} {2017})\BibitemShut {NoStop}%
\bibitem [{\citenamefont {Huthnance}(1975)}]{huthnance1975trapped}%
  \BibitemOpen
  \bibfield  {author} {\bibinfo {author} {\bibfnamefont {J.}~\bibnamefont
  {Huthnance}},\ }\bibfield  {title} {\bibinfo {title} {On trapped waves over a
  continental shelf},\ }\href@noop {} {\bibfield  {journal} {\bibinfo
  {journal} {Journal of fluid mechanics}\ }\textbf {\bibinfo {volume} {69}},\
  \bibinfo {pages} {689} (\bibinfo {year} {1975})}\BibitemShut {NoStop}%
\bibitem [{\citenamefont {Hendershott}(1973)}]{hendershott1973ocean}%
  \BibitemOpen
  \bibfield  {author} {\bibinfo {author} {\bibfnamefont {M.~C.}\ \bibnamefont
  {Hendershott}},\ }\bibfield  {title} {\bibinfo {title} {Ocean tides},\
  }\href@noop {} {\bibfield  {journal} {\bibinfo  {journal} {Eos, Transactions
  American Geophysical Union}\ }\textbf {\bibinfo {volume} {54}},\ \bibinfo
  {pages} {76} (\bibinfo {year} {1973})}\BibitemShut {NoStop}%
\bibitem [{\citenamefont {Ball}(1967)}]{ball1967edge}%
  \BibitemOpen
  \bibfield  {author} {\bibinfo {author} {\bibfnamefont {F.}~\bibnamefont
  {Ball}},\ }\bibfield  {title} {\bibinfo {title} {Edge waves in an ocean of
  finite depth},\ }in\ \href@noop {} {\emph {\bibinfo {booktitle} {Deep Sea
  Research and Oceanographic Abstracts}}},\ Vol.~\bibinfo {volume} {14}\
  (\bibinfo {organization} {Elsevier},\ \bibinfo {year} {1967})\ pp.\ \bibinfo
  {pages} {79--88}\BibitemShut {NoStop}%
\bibitem [{\citenamefont {Reznik}\ and\ \citenamefont
  {Zeitlin}(2011)}]{reznik2011resonant}%
  \BibitemOpen
  \bibfield  {author} {\bibinfo {author} {\bibfnamefont {G.}~\bibnamefont
  {Reznik}}\ and\ \bibinfo {author} {\bibfnamefont {V.}~\bibnamefont
  {Zeitlin}},\ }\bibfield  {title} {\bibinfo {title} {Resonant excitation of
  trapped waves by poincar{\'e} waves in the coastal waveguides},\ }\href@noop
  {} {\bibfield  {journal} {\bibinfo  {journal} {Journal of fluid mechanics}\
  }\textbf {\bibinfo {volume} {673}},\ \bibinfo {pages} {349} (\bibinfo {year}
  {2011})}\BibitemShut {NoStop}%
\bibitem [{\citenamefont {Nakahara}(2003)}]{nakahara2003geometry}%
  \BibitemOpen
  \bibfield  {author} {\bibinfo {author} {\bibfnamefont {M.}~\bibnamefont
  {Nakahara}},\ }\href@noop {} {\emph {\bibinfo {title} {Geometry, topology and
  physics}}}\ (\bibinfo  {publisher} {CRC Press},\ \bibinfo {year}
  {2003})\BibitemShut {NoStop}%
\bibitem [{\citenamefont {Zworski}(2012)}]{Zworski}%
  \BibitemOpen
  \bibfield  {author} {\bibinfo {author} {\bibfnamefont {M.}~\bibnamefont
  {Zworski}},\ }\href@noop {} {\emph {\bibinfo {title} {Semiclassical
  Analysis}}},\ Vol.\ \bibinfo {volume} {138}\ (\bibinfo  {publisher} {American
  Mathematical Society, Graduate Studies in Mathematics},\ \bibinfo {year}
  {2012})\BibitemShut {NoStop}%
\bibitem [{\citenamefont {Littlejohn}\ and\ \citenamefont
  {Flynn}(1991)}]{littlejohn1991geometric}%
  \BibitemOpen
  \bibfield  {author} {\bibinfo {author} {\bibfnamefont {R.~G.}\ \bibnamefont
  {Littlejohn}}\ and\ \bibinfo {author} {\bibfnamefont {W.~G.}\ \bibnamefont
  {Flynn}},\ }\bibfield  {title} {\bibinfo {title} {Geometric phases in the
  asymptotic theory of coupled wave equations},\ }\href@noop {} {\bibfield
  {journal} {\bibinfo  {journal} {Physical Review A}\ }\textbf {\bibinfo
  {volume} {44}},\ \bibinfo {pages} {5239} (\bibinfo {year}
  {1991})}\BibitemShut {NoStop}%
\bibitem [{\citenamefont {Colin~de Verd{\`\i}{\`e}re}\ and\ \citenamefont
  {Saint-Raymond}(2020)}]{colin}%
  \BibitemOpen
  \bibfield  {author} {\bibinfo {author} {\bibfnamefont {Y.}~\bibnamefont
  {Colin~de Verd{\`\i}{\`e}re}}\ and\ \bibinfo {author} {\bibfnamefont
  {L.}~\bibnamefont {Saint-Raymond}},\ }\bibfield  {title} {\bibinfo {title}
  {Attractors for two-dimensional waves with homogeneous hamiltonians of degree
  0},\ }\href@noop {} {\bibfield  {journal} {\bibinfo  {journal}
  {Communications on Pure and Applied Mathematics}\ }\textbf {\bibinfo {volume}
  {73}},\ \bibinfo {pages} {421} (\bibinfo {year} {2020})}\BibitemShut
  {NoStop}%
\bibitem [{\citenamefont {Haldane}(1988)}]{haldane1988model}%
  \BibitemOpen
  \bibfield  {author} {\bibinfo {author} {\bibfnamefont {F.~D.~M.}\
  \bibnamefont {Haldane}},\ }\bibfield  {title} {\bibinfo {title} {Model for a
  quantum hall effect without landau levels: Condensed-matter realization of
  the" parity anomaly"},\ }\href@noop {} {\bibfield  {journal} {\bibinfo
  {journal} {Physical review letters}\ }\textbf {\bibinfo {volume} {61}},\
  \bibinfo {pages} {2015} (\bibinfo {year} {1988})}\BibitemShut {NoStop}%
\bibitem [{Note1()}]{Note1}%
  \BibitemOpen
  \bibinfo {note} {\textcolor {black}{Note that this critical latitude should
  be confused with critical layers that occur when waves propagate in the
  presence of a mean flow.}}\BibitemShut {Stop}%
\bibitem [{\citenamefont {Faure}\ and\ \citenamefont
  {Zhilinskii}(2000)}]{faure2000topological}%
  \BibitemOpen
  \bibfield  {author} {\bibinfo {author} {\bibfnamefont {F.}~\bibnamefont
  {Faure}}\ and\ \bibinfo {author} {\bibfnamefont {B.}~\bibnamefont
  {Zhilinskii}},\ }\bibfield  {title} {\bibinfo {title} {Topological chern
  indices in molecular spectra},\ }\href@noop {} {\bibfield  {journal}
  {\bibinfo  {journal} {Physical {R}eview {L}etters}\ }\textbf {\bibinfo
  {volume} {85}},\ \bibinfo {pages} {960} (\bibinfo {year} {2000})}\BibitemShut
  {NoStop}%
\bibitem [{\citenamefont {Perrot}\ \emph {et~al.}(2019)\citenamefont {Perrot},
  \citenamefont {Delplace},\ and\ \citenamefont
  {Venaille}}]{perrot2019topological}%
  \BibitemOpen
  \bibfield  {author} {\bibinfo {author} {\bibfnamefont {M.}~\bibnamefont
  {Perrot}}, \bibinfo {author} {\bibfnamefont {P.}~\bibnamefont {Delplace}},\
  and\ \bibinfo {author} {\bibfnamefont {A.}~\bibnamefont {Venaille}},\
  }\bibfield  {title} {\bibinfo {title} {Topological transition in stratified
  fluids},\ }\href@noop {} {\bibfield  {journal} {\bibinfo  {journal} {Nature
  Physics}\ }\textbf {\bibinfo {volume} {15}},\ \bibinfo {pages} {781}
  (\bibinfo {year} {2019})}\BibitemShut {NoStop}%
\bibitem [{\citenamefont {Parker}\ \emph
  {et~al.}(2020{\natexlab{a}})\citenamefont {Parker}, \citenamefont {Marston},
  \citenamefont {Tobias},\ and\ \citenamefont {Zhu}}]{parker2020topological}%
  \BibitemOpen
  \bibfield  {author} {\bibinfo {author} {\bibfnamefont {J.~B.}\ \bibnamefont
  {Parker}}, \bibinfo {author} {\bibfnamefont {J.}~\bibnamefont {Marston}},
  \bibinfo {author} {\bibfnamefont {S.}~\bibnamefont {Tobias}},\ and\ \bibinfo
  {author} {\bibfnamefont {Z.}~\bibnamefont {Zhu}},\ }\bibfield  {title}
  {\bibinfo {title} {Topological gaseous plasmon polariton in realistic
  plasma},\ }\href@noop {} {\bibfield  {journal} {\bibinfo  {journal} {Physical
  {R}eview {L}etters}\ }\textbf {\bibinfo {volume} {124}},\ \bibinfo {pages}
  {195001} (\bibinfo {year} {2020}{\natexlab{a}})}\BibitemShut {NoStop}%
\bibitem [{\citenamefont {Parker}\ \emph
  {et~al.}(2020{\natexlab{b}})\citenamefont {Parker}, \citenamefont {Burby},
  \citenamefont {Marston},\ and\ \citenamefont
  {Tobias}}]{parker2020nontrivialb}%
  \BibitemOpen
  \bibfield  {author} {\bibinfo {author} {\bibfnamefont {J.}~\bibnamefont
  {Parker}}, \bibinfo {author} {\bibfnamefont {J.}~\bibnamefont {Burby}},
  \bibinfo {author} {\bibfnamefont {J.}~\bibnamefont {Marston}},\ and\ \bibinfo
  {author} {\bibfnamefont {S.~M.}\ \bibnamefont {Tobias}},\ }\bibfield  {title}
  {\bibinfo {title} {Nontrivial topology in the continuous spectrum of a
  magnetized plasma},\ }\href@noop {} {\bibfield  {journal} {\bibinfo
  {journal} {Physical Review Research}\ }\textbf {\bibinfo {volume} {2}},\
  \bibinfo {pages} {033425} (\bibinfo {year} {2020}{\natexlab{b}})}\BibitemShut
  {NoStop}%
\bibitem [{\citenamefont {Shankar}\ \emph {et~al.}(2017)\citenamefont
  {Shankar}, \citenamefont {Bowick},\ and\ \citenamefont
  {Marchetti}}]{shankar2017topological}%
  \BibitemOpen
  \bibfield  {author} {\bibinfo {author} {\bibfnamefont {S.}~\bibnamefont
  {Shankar}}, \bibinfo {author} {\bibfnamefont {M.~J.}\ \bibnamefont
  {Bowick}},\ and\ \bibinfo {author} {\bibfnamefont {M.~C.}\ \bibnamefont
  {Marchetti}},\ }\bibfield  {title} {\bibinfo {title} {Topological sound and
  flocking on curved surfaces},\ }\href@noop {} {\bibfield  {journal} {\bibinfo
   {journal} {Physical Review X}\ }\textbf {\bibinfo {volume} {7}},\ \bibinfo
  {pages} {031039} (\bibinfo {year} {2017})}\BibitemShut {NoStop}%
\bibitem [{\citenamefont {Green}\ \emph {et~al.}(2020)\citenamefont {Green},
  \citenamefont {Armas}, \citenamefont {de~Boer},\ and\ \citenamefont
  {Giomi}}]{green2020topological}%
  \BibitemOpen
  \bibfield  {author} {\bibinfo {author} {\bibfnamefont {R.}~\bibnamefont
  {Green}}, \bibinfo {author} {\bibfnamefont {J.}~\bibnamefont {Armas}},
  \bibinfo {author} {\bibfnamefont {J.}~\bibnamefont {de~Boer}},\ and\ \bibinfo
  {author} {\bibfnamefont {L.}~\bibnamefont {Giomi}},\ }\bibfield  {title}
  {\bibinfo {title} {Topological waves in passive and active fluids on curved
  surfaces: a unified picture},\ }\href@noop {} {\bibfield  {journal} {\bibinfo
   {journal} {arXiv preprint arXiv:2011.12271}\ } (\bibinfo {year}
  {2020})}\BibitemShut {NoStop}%
\bibitem [{\citenamefont {Marciani}\ and\ \citenamefont
  {Delplace}(2020)}]{marciani20}%
  \BibitemOpen
  \bibfield  {author} {\bibinfo {author} {\bibfnamefont {M.}~\bibnamefont
  {Marciani}}\ and\ \bibinfo {author} {\bibfnamefont {P.}~\bibnamefont
  {Delplace}},\ }\bibfield  {title} {\bibinfo {title} {Chiral maxwell waves in
  continuous media from berry monopoles},\ }\href
  {https://doi.org/10.1103/PhysRevA.101.023827} {\bibfield  {journal} {\bibinfo
   {journal} {Phys. Rev. A}\ }\textbf {\bibinfo {volume} {101}},\ \bibinfo
  {pages} {023827} (\bibinfo {year} {2020})}\BibitemShut {NoStop}%
\bibitem [{\citenamefont {Burns}\ \emph {et~al.}(2020)\citenamefont {Burns},
  \citenamefont {Vasil}, \citenamefont {Oishi}, \citenamefont {Lecoanet},\ and\
  \citenamefont {Brown}}]{burns2020dedalus}%
  \BibitemOpen
  \bibfield  {author} {\bibinfo {author} {\bibfnamefont {K.~J.}\ \bibnamefont
  {Burns}}, \bibinfo {author} {\bibfnamefont {G.~M.}\ \bibnamefont {Vasil}},
  \bibinfo {author} {\bibfnamefont {J.~S.}\ \bibnamefont {Oishi}}, \bibinfo
  {author} {\bibfnamefont {D.}~\bibnamefont {Lecoanet}},\ and\ \bibinfo
  {author} {\bibfnamefont {B.~P.}\ \bibnamefont {Brown}},\ }\bibfield  {title}
  {\bibinfo {title} {Dedalus: A flexible framework for numerical simulations
  with spectral methods},\ }\href@noop {} {\bibfield  {journal} {\bibinfo
  {journal} {Physical Review Research}\ }\textbf {\bibinfo {volume} {2}},\
  \bibinfo {pages} {023068} (\bibinfo {year} {2020})}\BibitemShut {NoStop}%
\bibitem [{\citenamefont {Iga}(1995)}]{iga1995transition}%
  \BibitemOpen
  \bibfield  {author} {\bibinfo {author} {\bibfnamefont {K.}~\bibnamefont
  {Iga}},\ }\bibfield  {title} {\bibinfo {title} {Transition modes of rotating
  shallow water waves in a channel},\ }\href@noop {} {\bibfield  {journal}
  {\bibinfo  {journal} {Journal of {F}luid {M}echanics}\ }\textbf {\bibinfo
  {volume} {294}},\ \bibinfo {pages} {367} (\bibinfo {year}
  {1995})}\BibitemShut {NoStop}%
\bibitem [{\citenamefont
  {Longuet-Higgins}(1968{\natexlab{a}})}]{longuet1968double}%
  \BibitemOpen
  \bibfield  {author} {\bibinfo {author} {\bibfnamefont {M.}~\bibnamefont
  {Longuet-Higgins}},\ }\bibfield  {title} {\bibinfo {title} {Double kelvin
  waves with continuous depth profiles},\ }\href@noop {} {\bibfield  {journal}
  {\bibinfo  {journal} {Journal of Fluid Mechanics}\ }\textbf {\bibinfo
  {volume} {34}},\ \bibinfo {pages} {49} (\bibinfo {year}
  {1968}{\natexlab{a}})}\BibitemShut {NoStop}%
\bibitem [{\citenamefont
  {Longuet-Higgins}(1968{\natexlab{b}})}]{longuet1968trapping}%
  \BibitemOpen
  \bibfield  {author} {\bibinfo {author} {\bibfnamefont {M.}~\bibnamefont
  {Longuet-Higgins}},\ }\bibfield  {title} {\bibinfo {title} {On the trapping
  of waves along a discontinuity of depth in a rotating ocean},\ }\href@noop {}
  {\bibfield  {journal} {\bibinfo  {journal} {Journal of Fluid Mechanics}\
  }\textbf {\bibinfo {volume} {31}},\ \bibinfo {pages} {417} (\bibinfo {year}
  {1968}{\natexlab{b}})}\BibitemShut {NoStop}%
\bibitem [{\citenamefont {Ren}\ \emph {et~al.}(2021)\citenamefont {Ren},
  \citenamefont {Fan}, \citenamefont {Xia}, \citenamefont {Chen}, \citenamefont
  {Yang}, \citenamefont {Zhong},\ and\ \citenamefont {Zhang}}]{ren2021robust}%
  \BibitemOpen
  \bibfield  {author} {\bibinfo {author} {\bibfnamefont {C.}~\bibnamefont
  {Ren}}, \bibinfo {author} {\bibfnamefont {X.}~\bibnamefont {Fan}}, \bibinfo
  {author} {\bibfnamefont {Y.}~\bibnamefont {Xia}}, \bibinfo {author}
  {\bibfnamefont {T.}~\bibnamefont {Chen}}, \bibinfo {author} {\bibfnamefont
  {L.}~\bibnamefont {Yang}}, \bibinfo {author} {\bibfnamefont {J.-Q.}\
  \bibnamefont {Zhong}},\ and\ \bibinfo {author} {\bibfnamefont
  {H.}~\bibnamefont {Zhang}},\ }\bibfield  {title} {\bibinfo {title} {Robust
  propagation of internal coastal kelvin waves in complex domains},\
  }\href@noop {} {\bibfield  {journal} {\bibinfo  {journal} {arXiv preprint
  arXiv:2102.03545, in press for Physical Review F - Letter}\ } (\bibinfo
  {year} {2021})}\BibitemShut {NoStop}%
\bibitem [{\citenamefont {Graf}\ and\ \citenamefont
  {Shapiro}(2018)}]{grafbulkedge2018}%
  \BibitemOpen
  \bibfield  {author} {\bibinfo {author} {\bibfnamefont {G.~M.}\ \bibnamefont
  {Graf}}\ and\ \bibinfo {author} {\bibfnamefont {J.}~\bibnamefont {Shapiro}},\
  }\bibfield  {title} {\bibinfo {title} {The {Bulk}-{Edge} {Correspondence} for
  {Disordered} {Chiral} {Chains}},\ }\href
  {https://doi.org/10.1007/s00220-018-3247-0} {\bibfield  {journal} {\bibinfo
  {journal} {Communications in Mathematical Physics}\ }\textbf {\bibinfo
  {volume} {363}},\ \bibinfo {pages} {829} (\bibinfo {year}
  {2018})}\BibitemShut {NoStop}%
\bibitem [{\citenamefont {Prodan}\ and\ \citenamefont
  {Schulz-Baldes}(2016)}]{ProdanEmilSchultz}%
  \BibitemOpen
  \bibfield  {author} {\bibinfo {author} {\bibfnamefont {E.}~\bibnamefont
  {Prodan}}\ and\ \bibinfo {author} {\bibfnamefont {H.}~\bibnamefont
  {Schulz-Baldes}},\ }\href {https://doi.org/10.1007/978-3-319-29351-6} {\emph
  {\bibinfo {title} {Bulk and Boundary Invariants for Complex Topological
  Insulators: From K-Theory to Physics}}}\ (\bibinfo  {publisher} {Mathematical
  Physics Studies},\ \bibinfo {year} {2016})\BibitemShut {NoStop}%
\bibitem [{\citenamefont {Boyd}(2018)}]{boyd2018dynamics}%
  \BibitemOpen
  \bibfield  {author} {\bibinfo {author} {\bibfnamefont {J.~P.}\ \bibnamefont
  {Boyd}},\ }\href@noop {} {\emph {\bibinfo {title} {Dynamics of the equatorial
  ocean}}}\ (\bibinfo  {publisher} {Springer},\ \bibinfo {year}
  {2018})\BibitemShut {NoStop}%
\bibitem [{\citenamefont {Heimpel}\ \emph {et~al.}(2005)\citenamefont
  {Heimpel}, \citenamefont {Aurnou},\ and\ \citenamefont
  {Wicht}}]{heimpel2005simulation}%
  \BibitemOpen
  \bibfield  {author} {\bibinfo {author} {\bibfnamefont {M.}~\bibnamefont
  {Heimpel}}, \bibinfo {author} {\bibfnamefont {J.}~\bibnamefont {Aurnou}},\
  and\ \bibinfo {author} {\bibfnamefont {J.}~\bibnamefont {Wicht}},\ }\bibfield
   {title} {\bibinfo {title} {Simulation of equatorial and high-latitude jets
  on jupiter in a deep convection model},\ }\href@noop {} {\bibfield  {journal}
  {\bibinfo  {journal} {Nature}\ }\textbf {\bibinfo {volume} {438}},\ \bibinfo
  {pages} {193} (\bibinfo {year} {2005})}\BibitemShut {NoStop}%
\bibitem [{\citenamefont {Cabanes}\ \emph {et~al.}(2017)\citenamefont
  {Cabanes}, \citenamefont {Aurnou}, \citenamefont {Favier},\ and\
  \citenamefont {Le~Bars}}]{cabanes2017laboratory}%
  \BibitemOpen
  \bibfield  {author} {\bibinfo {author} {\bibfnamefont {S.}~\bibnamefont
  {Cabanes}}, \bibinfo {author} {\bibfnamefont {J.}~\bibnamefont {Aurnou}},
  \bibinfo {author} {\bibfnamefont {B.}~\bibnamefont {Favier}},\ and\ \bibinfo
  {author} {\bibfnamefont {M.}~\bibnamefont {Le~Bars}},\ }\bibfield  {title}
  {\bibinfo {title} {A laboratory model for deep-seated jets on the gas
  giants},\ }\href@noop {} {\bibfield  {journal} {\bibinfo  {journal} {Nature
  Physics}\ }\textbf {\bibinfo {volume} {13}},\ \bibinfo {pages} {387}
  (\bibinfo {year} {2017})}\BibitemShut {NoStop}%
\bibitem [{\citenamefont {Deremble}\ \emph {et~al.}(2017)\citenamefont
  {Deremble}, \citenamefont {Johnson},\ and\ \citenamefont
  {Dewar}}]{deremble2017coupled}%
  \BibitemOpen
  \bibfield  {author} {\bibinfo {author} {\bibfnamefont {B.}~\bibnamefont
  {Deremble}}, \bibinfo {author} {\bibfnamefont {E.}~\bibnamefont {Johnson}},\
  and\ \bibinfo {author} {\bibfnamefont {W.}~\bibnamefont {Dewar}},\ }\bibfield
   {title} {\bibinfo {title} {A coupled model of interior balanced and boundary
  flow},\ }\href@noop {} {\bibfield  {journal} {\bibinfo  {journal} {Ocean
  Modelling}\ }\textbf {\bibinfo {volume} {119}},\ \bibinfo {pages} {1}
  (\bibinfo {year} {2017})}\BibitemShut {NoStop}%
\bibitem [{\citenamefont {Venaille}(2020)}]{venaille2020quasi}%
  \BibitemOpen
  \bibfield  {author} {\bibinfo {author} {\bibfnamefont {A.}~\bibnamefont
  {Venaille}},\ }\bibfield  {title} {\bibinfo {title} {Quasi-geostrophy against
  the wall},\ }\href@noop {} {\bibfield  {journal} {\bibinfo  {journal}
  {Journal of Fluid Mechanics}\ }\textbf {\bibinfo {volume} {894}} (\bibinfo
  {year} {2020})}\BibitemShut {NoStop}%
\bibitem [{\citenamefont {Fruchart}\ and\ \citenamefont
  {Carpentier}(2013)}]{fruchart2013introduction}%
  \BibitemOpen
  \bibfield  {author} {\bibinfo {author} {\bibfnamefont {M.}~\bibnamefont
  {Fruchart}}\ and\ \bibinfo {author} {\bibfnamefont {D.}~\bibnamefont
  {Carpentier}},\ }\bibfield  {title} {\bibinfo {title} {An introduction to
  topological insulators},\ }\href@noop {} {\bibfield  {journal} {\bibinfo
  {journal} {Comptes Rendus Physique}\ }\textbf {\bibinfo {volume} {14}},\
  \bibinfo {pages} {779} (\bibinfo {year} {2013})}\BibitemShut {NoStop}%
\bibitem [{\citenamefont {Bernevig}\ and\ \citenamefont
  {Hughes}(2013)}]{bernevig}%
  \BibitemOpen
  \bibfield  {author} {\bibinfo {author} {\bibfnamefont {B.~A.}\ \bibnamefont
  {Bernevig}}\ and\ \bibinfo {author} {\bibfnamefont {T.~L.}\ \bibnamefont
  {Hughes}},\ }\href@noop {} {\emph {\bibinfo {title} {Topological insulators
  and topological superconductors}}}\ (\bibinfo  {publisher} {Princeton
  university press},\ \bibinfo {year} {2013})\BibitemShut {NoStop}%
\bibitem [{\citenamefont {Dubrovin}\ \emph {et~al.}(1985)\citenamefont
  {Dubrovin}, \citenamefont {Fomenko},\ and\ \citenamefont
  {Novikov}}]{DubrovinBook}%
  \BibitemOpen
  \bibfield  {author} {\bibinfo {author} {\bibfnamefont {B.}~\bibnamefont
  {Dubrovin}}, \bibinfo {author} {\bibfnamefont {A.}~\bibnamefont {Fomenko}},\
  and\ \bibinfo {author} {\bibfnamefont {S.}~\bibnamefont {Novikov}},\
  }\href@noop {} {\emph {\bibinfo {title} {Modern Geometry--- Methods and
  Applications: Part II: The Geometry and Topology of Manifolds}}}\ (\bibinfo
  {publisher} {Springer Science and Business Media},\ \bibinfo {year}
  {1985})\BibitemShut {NoStop}%
\end{thebibliography}%

\end{document}